\documentstyle[preprint,aps,floats,fleqn,here]{revtex}

\ifx\nopictures Y \else \input epsf.tex \fi
\tighten
\draft
\begin{document}

\def\pAl{p_{A, l}}                               
\def\pBl{p_{B, l}}
\def\pal{p_{a, l}}                                 
\def\pbl{p_{b, l}}

\def\pAc{p_{A, cm}}                             
\def\pBc{p_{B, cm}}
\def\pac{p_{a, cm}}                             
\def\pbc{p_{b, cm}}

\def\pAh{p_{A, h}}                             
\def\pBh{p_{B, h}}
\def\pah{p_{a, h}}                             
\def\pbh{p_{b, h}}

\def\xh{\hat x}
\def\zh{\hat z}
\def\qTh{\hat q_T}
\def\sh{\hat s}
\def\th{\hat t}
\def\uh{\hat u}
\def \sigh {\hat sigma}

\def\beq{\begin{equation}}
\def\eeq{\end{equation}}
\def\bdm{\begin{displaymath}}
\def\edm{\end{displaymath}}
\def\bea{\begin{eqnarray}}
\def\eea{\end{eqnarray}}

\def\lp{\l^{\prime}}
\def\SeA{S_{eA}}                                        
\def\vg{\gamma^{*}}                                     
\def\alpi{\frac{\alpha_s}{\pi}}                         
\def\sFs{\frac{\sigma_0 F_l}{\SeA}}

\def\D0{D\0~}
\def\ov{\overline}
\def\ra{\rightarrow}
\def\cms{c.m. }

\setcounter{footnote}{1}
\renewcommand{\thefootnote}{\fnsymbol{footnote}}

\preprint{MSUHEP-90601}
\title{Semi-Inclusive Hadron Production at HERA: the Effect of
QCD Gluon Resummation}

\author{ P. Nadolsky, D. R. Stump, C.-P. Yuan}
\address{\rm Department of Physics and Astronomy, Michigan State University,\\
East Lansing, MI 48824, USA}

\date{\today}
\maketitle
\begin{abstract}
We present a formalism that improves  the applicability of
perturbative QCD in the current region of
semi-inclusive deep inelastic scattering. The formalism is based on
all-order resummation of large logarithms arising in the perturbative
treatment of hadron multiplicities and energy flows in this region.
It is shown that the current region of
semi-inclusive DIS is similar to the region of small transverse momenta
in  vector boson production at hadron colliders.
We use this resummation formalism to  describe
transverse energy flows and charged particle multiplicity
measured at the electron-proton collider HERA. We find good agreement
between our theoretical results and
experimental data for the transverse energy flows.

\end{abstract}
\newpage

\widetext

\section {Introduction}

It is well known that  perturbative Quantum Chromodynamics (pQCD) is a
very powerful but not omnipotent theory of the
strong interactions of elementary particles.
It can be successfully applied to the calculation of various physical
observables whenever the kinematics of the particle interaction
process implies the existence of a large scale $Q$ with
dimensionality of momentum.  For such a kinematic regime, the cross-section
of the hadronic process can be represented as a convolution of
the perturbatively calculable hard part, describing the
energetic short-range interactions of
hadronic constituents (partons), and several process-independent
nonperturbative functions, relevant to
the complicated strong dynamics at large distances.

The factorization of the hard and soft parts has proven to be a powerful
method for the calculation of  hadronic scattering cross-sections.
Unfortunately, near the boundaries of the kinematic phase space
the convergence of the perturbative solution can
be spoiled by the presence of large logarithms $\log r$, where $r$ is some
dimensionless function of the kinematic parameters of the system.
For instance, $r$ might be a small ratio of two momentum scales $P_1$ and
$P_2$ of the system, $r=P_1 / P_2$.

To handle this situation, techniques for the all-order resummation of
the logarithmically divergent terms have been developed
\cite{RT1,RT2,RT3,epem,CSS,BFKL}. These
techniques have been successfully used to improve the applicability of
perturbative QCD in several processes (calculation of energy correlations
in $e^+ e^-$ annihilation \cite{epem},
transverse momentum distributions in vector
boson \cite{CSS,DWS,LY,BY},di-photon \cite{diphoton}
and Higgs \cite{Higgs} production at
hadron colliders).

In this paper, we will  consider another process, the production
of hadrons in  deep-inelastic lepton-hadron scattering (DIS). As will
be discussed below, some features of this process are similar to the
$e^+ e^-$ hadroproduction and vector boson production, so that in
certain kinematic regions the description of this process requires
 all-order  resummation of the large logs which would otherwise
spoil the convergence of
the perturbative calculation.

Deep-inelastic lepton-hadron scattering (DIS) at large
momentum transfer $Q$ is one of the cornerstone processes to test pQCD.
Traditionally, the experimental study of
the fully inclusive DIS process, $e + A \ra e + X$ where $A$ is
usually a nucleon, and $X$ is any final state, is used to obtain
information about the Parton Distribution Functions of the nucleon (PDFs).
These functions
describe the long-range dynamics of  hadron interactions, and
are required by many pQCD calculations.

During the 1990's,  significant attention has also been paid to various
aspects of semi-inclusive deep inelastic scattering
(semi-inclusive DIS), for instance, the
semi-inclusive production of hadrons and jets, $e + A \ra e + B + X$ and
$e + A \ra e + jets + X$. In particular, the H1 and ZEUS
collaborations at HERA, and the E665 experiment at Fermilab performed
extensive experimental studies of the charged particle
multiplicity \cite{HERAchgd,HERAchgd2,E665} and hadronic transverse
energy flows \cite{H1z} at large momentum transfer $Q$. It was
found that some aspects of the data, e.g., the Feynman $x$
distributions, can be successfully explained in the framework of
perturbative QCD analysis \cite{Graudenz}. On the other hand,
the applicability of pQCD for the description of other features of
the process is limited.
For example, the perturbative calculation in  lowest orders
fails to describe the pseudorapidity or transverse momentum
distributions
of the final hadrons.  Under certain kinematic conditions the whole
perturbative expansion as a series in the QCD coupling may fail due to the
large logarithms mentioned earlier.

To be more specific, consider  semi-inclusive DIS production of
hadrons of a type~$B$. At large energies, we can neglect the masses
of  the participating particles.
%
%
In  semi-inclusive DIS at given energies of
the beams, any event can be characterized by two energy scales: the
virtuality of the exchanged photon $Q$ and the scale $q_T$ related to the
transverse momentum of the final hadron $B$.
The exact definition of $q_T$ will be
given in the main part of the paper. One may try to use pQCD in any of
three regions, where $Q$, $q_T$, or both $Q$ and $q_T$ are large.
The renormalization and factorization scales should be chosen to be
of the order of the large physical scale of the process.
In the limit $Q \ll q_T$ (photoproduction region) pQCD may
fail due to the large terms  $(\log Q/q_T)^n$ as\ $Q\ra 0$,
which should be resummed
into the parton distribution function of the virtual photon \cite{Kramer}.
The limit $Q \gg q_T$ is similar to the limit of a small transverse
momentum in  vector boson production where  logarithms of the
type $(\log q_T/Q)^n$  should be resummed in order to get a finite
cross-section of the process \cite{CSS}. Finally, in the
region $ q_T \approx Q$ one may
encounter another type of large logarithms corresponding to events
with large rapidity gaps $\Delta y$.
This type of large logarithms can be resummed with the help of the
Balitsky-Fadin-Kuraev-Lipatov (BFKL) formalism\cite{BFKL}.

The purpose of this paper is to discuss the resummation of large
logarithms in semi-inclusive DIS hadroproduction $e + A \ra e+ B + X$
in the limit $q_T \ll Q$.  These are large logs arising from to
the emission of soft and collinear partons, which we resum using the
formalism of Collins, Soper, and Sterman (CSS)~\cite{CSS}.
Our calculations are
based on the works of Meng, Olness, and Soper ~\cite{Meng1,Meng2}, who
analyzed the resummation technique for a particular
 energy distribution function of the
semi-inclusive DIS process. This energy distribution function
 receives contributions from all
possible final state hadrons, and  does not depend
on the specifics of fragmentation in the final state.

In this paper we present a more
general formalism than the one developed in \cite{Meng1,Meng2},
which will also account for the final state fragmentation of the partons.
This formalism requires the knowledge of the fragmentation functions
(FFs) describing  nonperturbative fragmentation of final partons into
observed hadrons.
Correspondingly, this formalism can consider
a wider class of physical observables including particle
multiplicities.
  Our calculations will be done in
the next-to-leading order of  perturbative QCD.
As an example of a practical application of our formalism,
we compare our calculation with
the H1 data on the pseudorapidity distributions of the
transverse energy flow~\cite{H1z} in the $\vg p$ center-of-mass
frame.  We also present predictions
for charged particle multiplicity.
Another goal of this study is
to find in which regions of kinematic parameters
the CSS resummation formalism is sufficient to
describe the existing data, and in
which regions significant contributions from
other hadroproduction mechanisms,
such as the BFKL interactions \cite{BFKL}, higher order corrections including
multijet production with \cite{Kramer} or without \cite{Catani}
resolved photon contributions, or  photoproduction showering \cite{Jung},
cannot be ignored.

The outline of the paper is as follows.
In Section \ref{2} we define the kinematic variables for
the semi-inclusive DIS process and specify the coordinate frames which will
be used throughout the following discussion.
In Section \ref{3} we derive the resummed cross-section formulas.
In Section \ref{4}
we extend the results of  Section \ref{3} to obtain the
resummed energy flows.
In Section \ref{5} we describe the matching between the resummed and
perturbative cross-sections and energy flows. We also discuss 
kinematic corrections, which should be applied to the resummed
cross-section to account for fast contraction of the phase space of the
perturbative cross-section at $q_T\approx Q$.
In Section \ref{6} we describe the results of Monte Carlo calculations
for the resummed cross-sections and energy flows. We present the
comparison of our calculation with the existing energy flow data from
HERA. We also suggest how one can reanalyze  the existing
HERA and Fermilab-E665 data on  charged particle production
in order to adapt it for  unambiguous extraction of the
non-perturbative Sudakov factors.

\section {Kinematic Variables \label{2}}

We follow notations which are similar to the ones used in \cite{Meng1,Meng2}.
In this Section we summarize them.

\begin{figure}[t]
\epsfysize 10cm
\epsffile{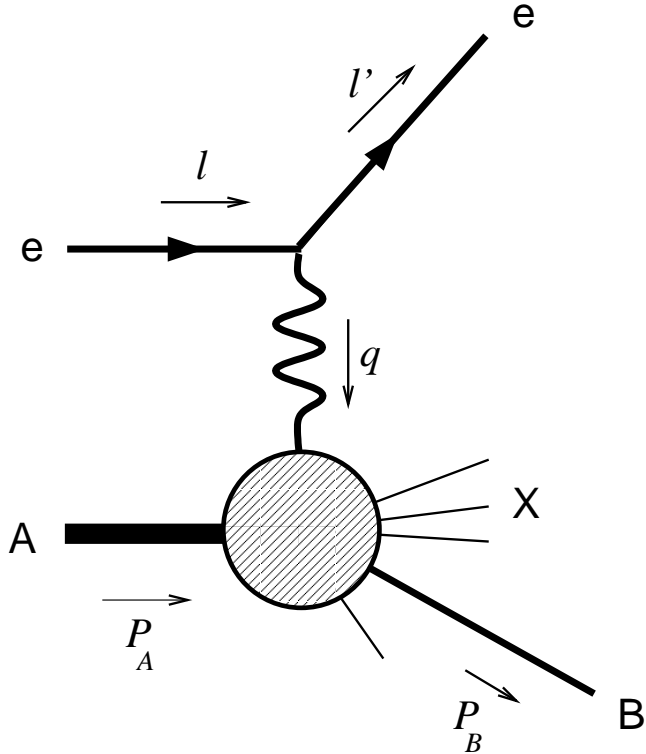}
\caption{\label{sDISeps} Semi-inclusive deep inelastic scattering}
\end{figure}

We limit the discussion to the case of semi-inclusive DIS at the
$ep$ collider HERA.
We consider the process
\begin{equation}
e+A\rightarrow e+B+X,   \label{sdis2}
\end{equation}
where $e$ is an electron or positron, $A$ is a proton,
$B$ is a hadron observed in the final state, and $X$ represents
any other particles in the final state in the sense of inclusive
scattering (Fig.~\ref{sDISeps}).
We denote the momenta of $A$ and $B$ by $p_{A}^{\mu}$ and
$p_{B}^{\mu}$, and the momenta of the lepton in the initial
and final states by $l^{\mu}$ and $l^{\prime\mu}$.
Also, $q^{\mu}$ is the momentum transfer to the hadron system,
$q^{\mu}=l^{\mu}-{\lp}^{\mu}$. For most of this paper, until the
discussion of  charged particle multiplicity, we
will neglect the particle masses.

We assume that the initial lepton and hadron interact only through a single
photon exchange.
Therefore, $q^{\mu}$ also has the
meaning of the 4-momentum of the exchanged virtual photon $\vg$.

\subsection{Lorentz scalars}
For further discussion, we define five Lorentz scalars relevant to the
process (\ref{sdis2}). The first is the center-of-mass energy of the
initial hadron and lepton $\sqrt{\SeA}$ where
\begin{equation} S_{eA}=(p_{A}+l)^{2}=2 p_{A}\cdot l.      \label{SeA}
\end{equation}
We will also use the conventional DIS variables $x$ and $Q^{2}$ which are
defined from the momentum transfer $q^{\mu}$ by
\begin{equation}
Q^{2}=-q^{2}=2\ell\cdot\ell^{\prime},
\end{equation}
\begin{equation}
x=\frac{Q^{2}}{2p_{A}\cdot{q}}\ .
\label{x}
\end{equation}
In principle, $x$ and $Q^{2}$ can be completely  determined
in an experimental event  by measuring the momentum of the outgoing lepton.

Next we define a scalar $z$ related to the momentum of the
final hadron  state $B$ by
\begin{equation}
\label{z}
z=\frac{p_{B}\cdot p_{A}}{q\cdot p_{A}}=\frac{2x p_{B}\cdot p_{A}}{Q^{2}}~.
\end{equation}
The variable $z$ plays an important role in the description of
fragmentation in the final state. In particular, in the quark-parton
model (or in the leading order perturbative calculation) it is equal
to the fraction of the fragmenting parton's momentum carried away by
the observed hadron.

The next relativistic invariant $q_{T}^2$ is the square of the component
of the virtual photon's 4-momentum $q^{\mu}$ that is transverse to the
4-momenta of  the initial and final hadrons:
\begin{equation}
q_{T}^{2}=-q_{t}^{\mu}q_{t\mu},
\end{equation}
where
\begin{equation}
q_{t}^{\mu}=q^{\mu}-p_{A}^{\mu}\frac{q\cdot p_{B}}{p_{A}\cdot p_{B}}
-p_{B}^{\mu}\frac{q\cdot p_{A}}{p_{A}\cdot p_{B}}. \label{qTmu}
\end{equation}
The orthogonality of  $q_{t}^{\mu}$ to both $p_{A}^{\mu}$
and $p_{B}^{\mu}$, that is $q_{t}\cdot p_{A}=q_{t}\cdot p_{B}=0$, follows
immediately from its definition (\ref{qTmu}).

The variable $q_T$  plays the same
role  in the semi-inclusive DIS resummation
as the transverse momentum of a vector boson in
resummation of vector boson production at hadron colliders.
In particular, the theoretical cross-section calculated in
a fixed order of pQCD is divergent
in the limit $q_T \ra 0$, so that all-order resummation
is needed to make the predictions of the theory finite
in this limit.

\begin{figure}[tb]
\epsfysize 10cm
\epsffile{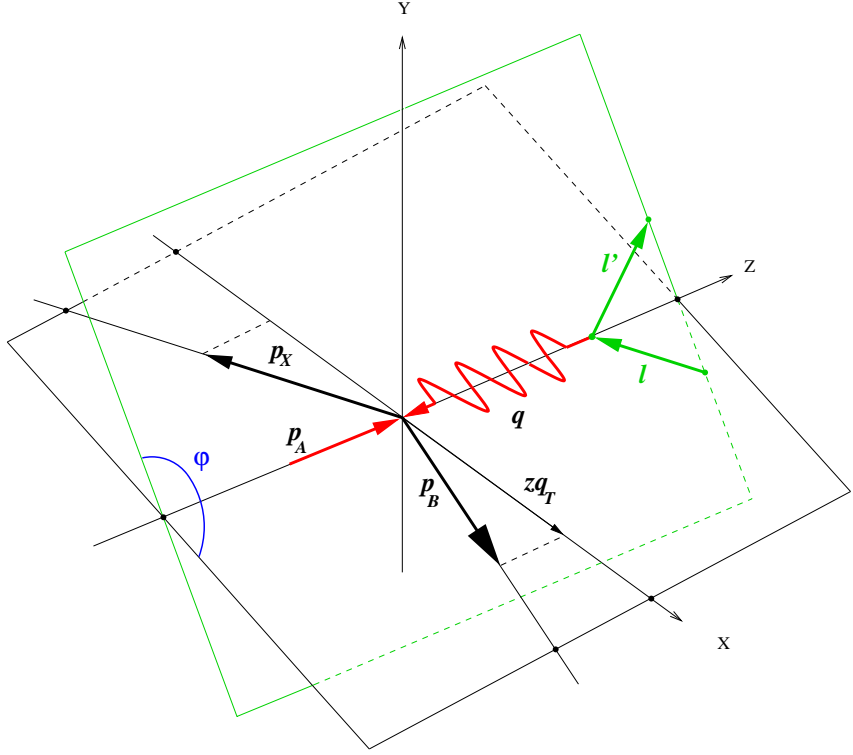}
\caption{Geometry of the particle momenta in the hadron frame \label{h}}
\end{figure}

In the analysis of kinematics, we will use two
reference frames. The first is
the center-of-mass ({\it c.m.})
frame of the initial hadron and the virtual photon.
The second is a special type of Breit frame which we will call,
depending on whether the initial state is a hadron or a parton,  the
{\it hadron} or {\it parton} frame.
As was shown in \cite{Meng1,Meng2},
by using the hadron frame one can organize
the resummation formalism for semi-inclusive DIS in a way that is
similar to the case of vector boson production.
On the other hand, many experimental results are presented in the $\vg
p$ c.m.~frame.
We will use a subscript $h$ and
$cm$ to denote kinematic variables in the hadron or \cms frame.

\subsection{Hadron frame}
Following Meng et al. \cite{Meng1,Meng2} the hadron frame is defined by
two conditions: (a) the 4-momentum of the virtual photon is purely
space-like, and (b) the momentum of the outgoing hadron $B$ lies in the $xz$
plane. The directions of
particle momenta in this frame are shown in Fig.~\ref{h}.

In this frame the proton $A$ moves in the $+z$
direction, while the momentum transfer $\vec{q}$ is in
the $-z$ direction, and $q^{0}$ is 0:
\begin{equation}
q^{\mu}_h=(0,0,0,-Q),
\end{equation}
\begin{equation}
p_{A, h}^{\mu}=\frac{Q}{2x}\Bigl( 1,0,0,1 \Bigr).
\end{equation}
The momentum of the final-state hadron $B$ is 
\begin{equation}
p_{B, h}^{\mu}=\frac{zQ}{2}
\Bigl(1+\frac{q_{T}^{2}}{Q^{2}},\frac{2 q_{T}}{Q},
0,\frac{q_{T}^{2}}{Q^{2}}-1 \Bigr).
\end{equation}
The incoming and outgoing
lepton momenta in the hadron frame are defined in terms of
variables $\psi$ and $\phi$ as follows \cite{phipsi}:
\begin{equation}
\ell^{\mu}_{h}=\frac{Q}{2}(\cosh\psi,\sinh\psi\cos\phi,\sinh\psi\sin\phi,-1),
\label{lh}
\end{equation}
\begin{equation}
\ell^{\prime\mu}_{h}
=\frac{Q}{2}(\cosh\psi,\sinh\psi\cos\phi,\sinh\psi\sin\phi,+1) .
\end{equation}
Note that $\phi$ is the azimuthal angle of $\vec{\ell}_h$
or $\vec{\ell}^{\prime}_h$ around the $z$-axis. 
$\psi$  is a parameter of a boost which relates the hadron frame to a 
lepton Breit frame in which $\ell^\mu = (Q/2, 0, 0, -Q/2)$. 
By (\ref{SeA}) and
(\ref{lh}) we find that
\begin{equation}
\cosh\psi=\frac{2x  \SeA}{Q^{2}}-1 = \frac{2}{y} -1,
\end{equation}
where the conventional DIS variable $y$ is defined as
\beq
y =\frac{Q^2}{x\SeA}.
\eeq
The allowed range of the variable $y$ in  deep-inelastic
scattering is $0\leq y \leq 1$; therefore $\psi\geq 0$.

The transverse part of the virtual photon momentum $q_{t}^{\mu}$
has a simple form in
the hadron frame; it can be shown that \begin{equation}
q_{t, h}^{\mu}=(-\frac{q_{T}^{2}}{Q},-q_{T},0,-\frac{q_{T}^{2}}{Q})~.
\end{equation}
In other words, $q_T$ is the magnitude of the transverse component of
$\vec q_{t, h}$.
The transverse momentum $p_T$ of the final state hadron $B$ in this
frame is simply related to $q_T$, by
\beq p_T = z q_T.  \eeq
Also, the pseudorapidity of $B$ in the hadron frame is
\beq
\eta_h \equiv -\log \Bigl(\tan\frac{\theta_{B, h}}{2}\Bigr) =\log
\frac{q_T}{Q}.  \eeq

The resummed  cross-section will be
derived using the hadron frame.
To transform the result to other frames, it is useful to express
the basis vectors of the hadron frame ($T^\mu, \ X^\mu,\ Y^\mu, \
Z^\mu$) in terms of the particle momenta \cite{Meng1}.
For an arbitrary coordinate frame,
\bea
T^\mu &=& \frac{q^\mu + 2 x p_A^\mu}{Q},
\nonumber \\
X^\mu &=&\frac{1}{q_T}\Biggl(\frac{p^\mu_B}{z} - q^\mu - \Bigl[ 1+
\frac{q_T^2}{Q^2}\Bigr] x p_A^\mu
\Biggr),
\nonumber \\
Y^\mu &=& \epsilon^{\mu\nu\rho\sigma} Z_\nu T_\rho X_\sigma,
\nonumber \\
Z^\mu &=& -\frac{q^\mu}{Q}.
\eea
If these relations are evaluated in the hadron frame, the basis
vectors $T^\mu, \ X^\mu,\ Y^\mu, \
Z^\mu$ are $(1,0,0,0),\ (0,1,0,0), (0,0,1,0), (0,0,0,1)$,
respectively.

The relationships
between the hadron-frame variables and the HERA lab-frame momenta are
presented in Appendix A.

\subsection{Photon-hadron center-of-mass frame}

\begin{figure}[h]
\epsfysize 10cm
\epsffile{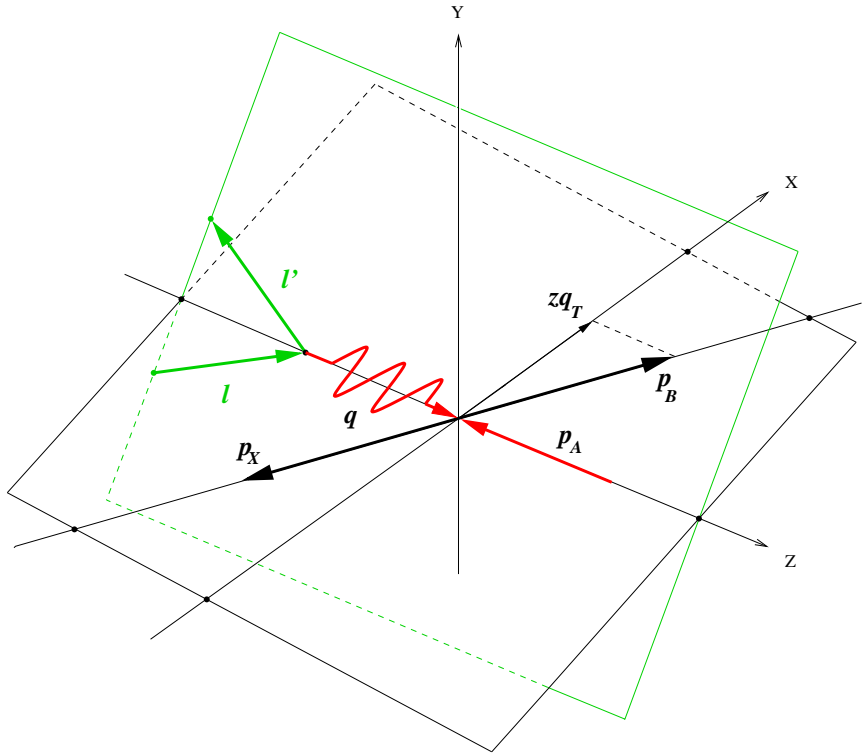}
\caption{ Particle momenta in the $\vg p$ c.m. frame \label{cm}}
\end{figure}

The center-of-mass frame of the proton $A$ and virtual photon $\vg$ is
defined by the condition $\vec \pAc + \vec q_{cm} =0$. The relationship
between particle momenta in this frame is illustrated in Fig.~\ref{cm}.
As in the hadron frame, the momenta
$\vec q_{cm}$ and  $\vec \pAc$ in the \cms frame are directed along the
$z$ axis.
The coordinate transformation from the hadron frame into the $\vg p$
\cms   frame consists of (a) a boost in the direction of the virtual photon
and (b) inversion of the direction of the $z$ axis, which is needed to
make the definition of the \cms   frame consistent with the one adopted in
HERA experimental publications. In the \cms   frame the momentum of
$\vg$ is
\beq
q_{cm}^\mu = \Biggl(
\frac{W^2-Q^2}{2 W}, 0 , 0,\frac{W^2+Q^2}{2 W}\Biggr),
\eeq
where $W$ is
the \cms   energy of the $\vg p$ collisions,
\beq
 W^2=(p_A+q)^2.
\eeq

The momenta of the initial and final hadrons $A$ and $B$ are given by
\beq
p^\mu_{A,cm} = \Biggl(
\frac{W^2+Q^2}{2 W}, 0 , 0,- \frac{W^2+Q^2}{2 W}\Biggr),
\label{pAc}
\eeq
\beq
p^\mu_{B,cm} = \Biggl( E_B, \pBc \sin \theta^{cm}_B, 0,
\pBc \cos \theta^{cm}_B \Biggr).
\label{pBc}
\eeq

Since the hadron and c.m.~frames are related by a boost along the
$z$-direction, the expression for the transverse momentum of the final
hadron $B$ in the \cms   frame is the same as the one in the hadron frame,
\beq   \label{pTB}
p_T = z q_T.
\eeq
Also, similar to the case of the hadron frame, the relationship between
$q_T$ and the pseudorapidity of $B$ in the \cms   frame is  simple,
\beq
\label{qTetacm}
 q_T = W e^{-\eta_{cm}}.
\eeq
The limit of small $q_T$, which is most relevant for our resummation  
calculation,
corresponds to the region of large pseudorapidities
in the hadronic c.m. frame. Since in this case the final parton
is produced in the direction of the momentum of the virtual photon, 
the region of large
$\eta^{cm}$ is also called the {\it current region}.

\subsection
{Parton kinematics}

The kinematic variables and momenta discussed so far are all laboratory
variables. Next, we relate these to parton variables.

Let $a$ denote the parton in $A$ that participates in the hard scattering,
with momentum $p_{a}^{\mu}=\xi_{a}p_{A}^{\mu}$.
Let $b$ denote the parton of which $B$ is a fragment,
with momentum $p_{b}^{\mu}=p_{B}^{\mu}/\xi_{b}$.
The momentum fractions $\xi_{a}$ and $\xi_{b}$ range from
0 to 1. At the parton level, we introduce the Lorentz scalars
$\xh,\ \zh, \ \hat q_T$  analogous to the ones at the hadron level:
\beq
\xh =\frac{Q^{2}}{2p_{a}\cdot{q}}= \frac{x}{\xi_a},
\eeq
\beq
\zh=\frac{p_{b}\cdot p_{a}}{q\cdot p_{a}}=\frac{z}{\xi_b},
\eeq
\beq
\qTh^{2} = -\hat q_t^\mu \hat {q_t}_\mu.
\eeq
Here $\hat q_T^\mu$ is the component of $q^\mu$ which is
orthogonal to the parton 4-momenta $p^\mu_a$ and $p^\mu_b$,
\bdm
 \hat q_t\cdot p_a = \hat q_t \cdot p_b =0.
\edm
Therefore,
\beq
\hat q_{t}^{\mu}=q^{\mu}-p_{a}^{\mu}\frac{q\cdot p_{b}}{p_{a}\cdot p_{b}}
-p_{b}^{\mu}\frac{q\cdot p_{a}}{p_{a}\cdot p_{b}}. \label{qTmup}
\eeq
In the case of massless initial and final hadrons   the hadronic and
partonic $q_T$'s coincide,
\beq
\qTh =q_T.
\eeq

\newpage
\begin{figure}[H]
\epsfysize 18cm
\epsffile{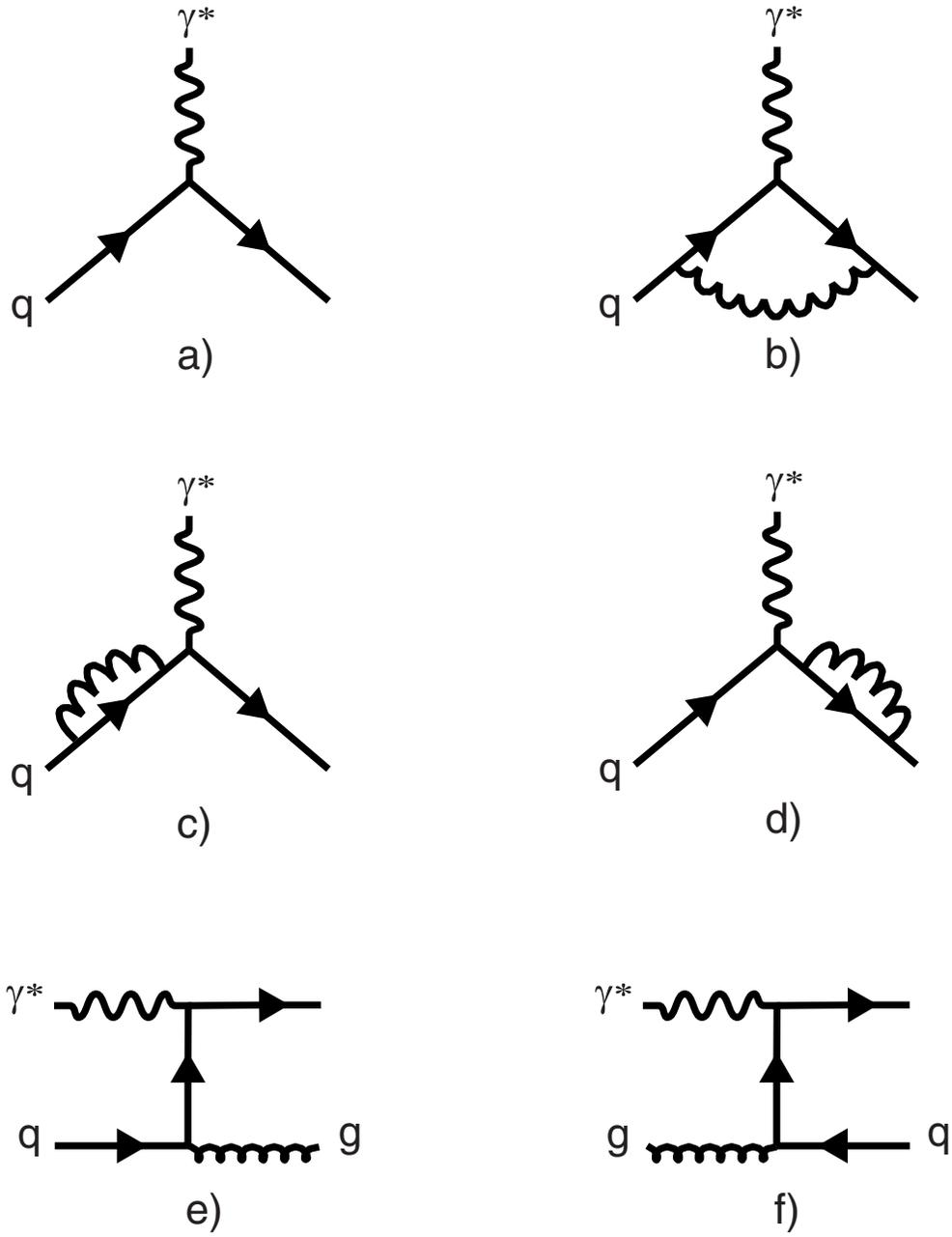}
\caption{\label{diags} Feynman diagrams for  semi-inclusive DIS:  a)
LO; b-d)NLO virtual diagrams; e-f) NLO real emission diagrams}
\end{figure}
\newpage
\section
{The resummed NLO cross-section \label{3}}

The knowledge of five Lorentz scalars
$\SeA,\ Q,\ q_T, \ x, \ z$ and the lepton azimuthal angle
$\phi$ in the hadron
frame is sufficient to specify unambiguously the kinematics of the
semi-inclusive scattering event $ e+ A \ra e+ B +X$.
In the following, we will discuss the hadron
cross-section $d \sigma_{BA}$,
which
is related to the parton cross-section $d \hat \sigma_{ba}$ by
\begin{equation}
\frac{d\sigma_{BA}}{dx dz dQ^{2} dq_{T}^{2} d\phi}=\sum_{a,b}
\int_{z}^{1}
\frac{d\xi_{b}}{\xi_{b}} D_{B/b}(\xi_{b}, \mu_D)\int_{x}^{1}
\frac{d\xi_{a}}{\xi_{a}}
F_{a/A}(\xi_{a}, \mu_F)
\frac{d\hat \sigma_{ba}}{d\hat{x} d\hat{z} dQ^{2} dq_{T}^{2} d\phi}.
\label{hadcs}
\end{equation}
Here
$F_{a/A}(\xi_a,\mu_F)$ denotes the distribution function (PDF)
of the parton of a
type $a$ in the hadron $A$, and
$D_{B/b}(\xi_b,\mu_D)$ is the fragmentation
function  (FF) for  parton type $b$ and
final hadron $B$. The parameters
$\mu_F$ and $\mu_D$ are the factorization scales for
the PDFs and FFs.
In the following discussion and calculations, we assume that these
factorization scales and the renormalization scale $\mu_R$
are the same,
\bdm
\mu_F =\mu_D =\mu_R \equiv \mu.
\edm

The analysis of semi-inclusive DIS can be conveniently organized by  
separating the dependence of the parton and hadron
cross-sections on the leptonic angle $\phi$ and the boost parameter $\psi$
from the other kinematic variables $x,\ z,\ Q$  and $q_T$ \cite{phipsi}.
Following \cite{Meng1}, we express
the hadron (or parton) cross-section 
as a sum over
products of functions of these lepton angles in the hadron frame
$A_{\alpha}(\psi, \phi)$, and structure functions
$V^{(\alpha)}_{BA}(x , z , Q^2, q_T^2)$
(or $\hat V^{(\alpha)}_{ba}(\xh ,\zh,Q^2,q_T^2)$, respectively):
\beq
 \frac{d \sigma_{BA}}{d x d z  dQ^{2}  dq_T^2 d\phi}=
\sum_{\alpha=1}^{4} V^{(\alpha)}_{BA}( x , z, Q^2, q_T^2)
A_{\alpha}(\psi,\phi), \label{ang1}
\eeq
\beq
 \frac{d \hat \sigma_{ba}}{d\xh d\zh  dQ^{2}  dq_T^2 d\phi}=
\sum_{\alpha=1}^{4} \hat V^{(\alpha)}_{ba}(\xh,\zh, Q^2, q_T^2)
A_{\alpha}(\psi,\phi). \label{ang2}
\eeq

At the energy of HERA, hadroproduction via parity-violating
$Z$-boson exchanges can be neglected,
  and only four out of the nine angular functions
listed  in \cite{Meng1}   contribute to the cross-sections
 (\ref{ang1}-\ref{ang2}). They are
\bea
 A_1 &=& 1+ \cosh^2 \psi,    \nonumber              \\
 A_2 &=& -2,                  \nonumber              \\
 A_3 &=& - \cos \phi \sinh 2\psi, \nonumber \\
 A_4 &=& \cos 2 \phi \sinh^2 \psi.
\eea
We will
assume that the angle $\phi$ is not monitored in the experiment, so that it
will be  integrated out in the following discussion. Correspondingly,
our numerical result for $d\sigma/(dx dz dQ^2 dq_T^2)$ will not depend on
terms in (\ref{ang1}-\ref{ang2}) proportional to the angular functions
  $A_3$ and $A_4$, which integrate to zero.

Out of the four structure functions,
only $\hat V^{(1)}_{ba}$
receives contributions from both
leading and next-to-leading order diagrams. Also, only
the $\hat V^{(1)}_{ba}$ structure function diverges in the limit $q_T \ra
0$.

The leading order (LO) parton process is $e+a\rightarrow e+b$
where the initial parton $a$ is a quark from the proton
and the final quark $b$ (which is the same as $a$) fragments
into the hadron $B$. The Feynman diagram for this process is shown in
Fig.~\ref{diags}a.
There is no LO contribution to semi-inclusive DIS
from gluons.

The next-to-leading order (NLO) corrections  are shown in
Figs.~\ref{diags}b-f.  At this order, we need to account for the virtual
corrections to the LO subprocess $\stackrel{(-)}{q}\vg \ra
\stackrel{(-)}{q}$ (Figs.~\ref{diags}b-d), as well as for the diagrams
describing the subprocesses $\stackrel{(-)}{q}\vg \ra \stackrel{(-)}{q}g$ and
$g\vg\ra q\bar q$, with the subsequent fragmentation of the final-state
quark, antiquark or gluon (Figs.~\ref{diags}e-f).

Conservation of total 4-momentum in the real
emission subprocesses (Figs.~\ref{diags}e-f)
allows us to write the
momentum  of the unobserved final state parton (e.g.
the gluon in Fig.~\ref{diags}e) as
\begin{equation}\label{pg}
p_{g}^{\mu}=q^{\mu}+p_{a}^{\mu}-p_{b}^{\mu}.
\end{equation}
When there is
no gluon radiation ($p_g^\mu= 0$) the momentum of $b$ is
$p_{b}^\mu=p_{a}^\mu +q^\mu$,
and, according to (\ref{qTmup}), $q_{T}^2=-q_{t}\cdot q_{t}=0$.
Thus, a non-zero $q_T$ in the event is an effect of gluon radiation.
In the
region $q_T \ra 0$, either softness or collinearity of the unobserved partons
will create infrared singularities, which make the perturbative
result unreliable. The sum of the real and virtual diagrams is made
finite by order-by-order cancellation of the soft singularities arising
from the real and virtual pieces, and by absorption of the collinear
singularities into the parton distribution and fragmentation functions.
Nonetheless, this cancellation does not guarantee rapid convergence of the
perturbative calculation, which will typically contain large logarithms
$\log q_T/Q$ countering the smallness of
the strong coupling.

The slow convergence of the perturbative series at $q_T \ra 0$ can be
corrected by  resummation of the most singular logarithmic terms.
It is done in the following way. First, we extract the terms in the 
squared amplitudes of the
real emission diagrams Figs.~\ref{diags}e-f
that are most singular in the limit $q_T \ra 0$; we refer to these terms
as the {\it asymptotic piece}. These terms are proportional to $1/q_T^2$ and,
as it was mentioned above, they 
appear only in the $\hat V^{(1)}_{ba}$ structure
function.  Thus, the structure function $\hat V^{(1)}_{ba} $ is represented
as
\beq \label{V1aY} \hat V^{(1)}_{ba}(\xh,\zh,Q^2, q_T^2) = \biggl( \hat
V^{(1)}_{ba}(\xh,\zh,Q, q_T^2)\biggr )_{asym} + \hat
Y^{(1)}_{ba}(\xh,\zh,Q^2,q_T^2),
\eeq
where $(\hat V^{(1)}_{ba})_{asym}$ is
${\cal O} (1/q_T^2)$, and $\hat Y^{(1)}_{ba}$ is finite in the limit $q_T
\ra 0$.  The 
asymptotic piece of  the NLO hadron cross-section (\ref{hadcs}) is
\bea
& &\Biggl(\frac{d \sigma_{BA}}{ d x d z d Q^2 d
q_T^2 d\phi} \Biggr)_{asym}=
\sFs \alpi \frac{e^2}{2 q_T^2}  \frac{A_1(\psi, \phi)}{2 \pi}
\nonumber \\
&\times& \sum_{j}
e_j^2 \Biggl[ D_{B/j}(z,\mu)\Bigl\{ (P_{qq} \circ f_{j/A})(x,\mu)
+(P_{qg}\circ f_{g/A})(x,\mu)\Bigr\}
\nonumber \\
&+& \Bigl\{ (D_{B/j}\circ
P_{qq})(z,\mu) +( D_{B/g}\circ P_{gq})(z,\mu)\Bigr\} f_{j/A}(x,\mu)
\nonumber \\
&+& 2 D_{B/j}(z,\mu) f_{j/A}(x,\mu)  \Bigl\{C_F \log
\frac{Q^2}{q_T^2}-\frac{3}{2}C_F\Bigr\} +
{\cal
  O}\Bigl(\biggl(\frac{\alpha_s}{\pi}\biggr)^2, q_T^2\Bigr)\Biggr].
\label{asym}
\eea
Here $e e_j$ is the electric
charge of the participating quark or antiquark of flavor $j$;
the parameter $\sigma_0$ is
\beq
\sigma_0 \equiv \frac{Q^2}{ 4 \pi \SeA x^2} \Bigl( \frac{e^2}{2} \Bigr) ;
\eeq
the factor $F_l$, that comes from the leptonic side, is defined by
\beq
F_l = \frac{e^2}{2} \frac{1}{Q^2};
\eeq
the color factor $C_F=(N_c^2-1)/(2
N_c )=4/3.$
The convolution in (\ref{asym}) is defined as
\begin{equation} (f \circ g)
(x,\mu)=\int_{x}^{1}f(x/\xi, \mu)g(\xi , \mu)\frac{d\xi}{\xi}~.
\end{equation}
The functions $P_{ij}(x)$ entering the convolution integrals in
(\ref{asym}) are the
familiar splitting kernels:
\beq
P_{qq}(x) = C_F \Biggl(\frac{1+ x^2}{ 1 - x } \Biggr)_+  ,
\eeq
\beq
P_{qg}(x) = \frac{1}{2} \biggl( 1 - 2 x +2 x^2 \biggr) ,
\eeq
\beq
P_{gq}(x) = C_F \frac{1 + (1- x)^2}{x} .
\eeq
The finite piece $Y^{(1)}_{BA}$ of the hadron cross-section
and the other structure
functions $V^{(i)}_{ba} \mbox{ for } \ i=2,3,4$ can be derived in a
straightforward way from the
expression for the cross-section of the real emission subprocesses,
which is
presented in Appendix B.  

Next, we use the perturbative asymptotic piece (\ref{asym})
to derive the
${\cal O}(\alpha_s)$ expression for the resummed cross-section
$(d \sigma / (d x d z dQ^2 dq_T^2 d\phi))_{resum}$. In the
Collins-Soper-Sterman resummation formalism \cite{CSS}, the cross-section
 is written as a Fourier
integral over a variable $\vec{b}$
conjugate to $\vec{q}_{T}$,
\begin{equation}
\label{resum}
\frac{d \sigma_{BA}}{d x d z d Q^2 d q_T^2 d \phi}=
\sFs \frac{ A_1 (\psi, \phi)}{2}
\int \frac{d^{n-2}b}{(2\pi)^{n-2}} e^{i\vec q_T \cdot \vec
b } W_{BA}(b, x, z , Q)+Y_{BA}.
\end{equation}
The term containing  the integral of $W_{BA}(b)$, which we name
the {\it CSS piece}, absorbs the asymptotic contributions from
all orders. The second term, which is the {\it finite piece}, has
the form
\beq
Y_{BA} = Y^{(1)}_{BA} + \sum_{\alpha=2}^{4} V^{(\alpha )}_{BA} A_{\alpha} (
\psi, \phi ).
\eeq

The Fourier transform is performed in the space of $n= 4
- \epsilon$ dimensions, in which  the asymptotic piece of the real emission
subprocesses generates terms which are proportional to $1/\epsilon$ 
and divergent as $\epsilon \rightarrow 0$.
Upon summation of the real and virtual diagrams, some of these
terms, specifically those corresponding to  soft singularities, cancel.
The remaining $1/\epsilon$ poles  correspond to collinear
singularities. They are later absorbed into the redefined NLO PDF or FF,
rendering a final expression that is nonsingular as $\epsilon
\rightarrow 0$.

According to \cite{CSS}, the form of the resummed
structure function $W_{BA}(b, x, z, Q)$ at small values of  $b$ is
\begin{equation}\label{W}
W_{BA}(b, x, z, Q)=\sum_j e_j^2
(D_{B/b} \circ {\cal C}^{out}_{bj})(z, b, \mu) ({\cal
C}_{ja}^{in} \circ F_{a/A}) (x, b, \mu)~e^{-S_{BA}(b, Q)}.
\end{equation}
In the limit of small $b$ and large $Q$, the Sudakov function
$S_{BA}(b, Q)$ does not depend on the types of the external hadrons
and looks like
\begin{equation} \label{sud_p}
S_{BA}(b , Q)=\int_{C_{1}^{2}/b^{2}}^{C_{2}^{2}Q^{2}}
\frac{d\ov \mu^{2}}{\ov \mu^{2}}\left(A(\alpha_s(\ov \mu), C_1)
\ln\frac{C_{2}^{2}Q^{2}}{\ov \mu^{2}} +B (\alpha_s(\ov \mu ),
C_1,C_2)\right),
\end{equation}
with
\beq A(\alpha_s(\ov \mu),
C_1)=\sum_{k=1}^{\infty} A_k (C_1) \Biggl(\frac{\alpha_s (\ov
\mu)}{\pi}\Biggr)^k,
\eeq
\beq B(\alpha_s(\ov \mu), C_1,
C_2)=\sum_{k=1}^{\infty} B_k (C_1, C_2) \Biggl(\frac{\alpha_s (\ov
\mu)}{\pi}\Biggr)^k.
\eeq
The integration in (\ref{sud_p}) is
performed between two scales of order $1/b$ and $Q$. The constants
$C_1$ and $C_2$ determining the exact integration range  are
{\it a priori} unknown, and their
variation allows us to test the scale invariance of the resummed
cross-section.  There are convenient choices of $C_1$, $C_2$ for which some
logarithms in $\hat W_{ba}(b,\xh,\zh, Q)$ cancel.

The functions ${\cal C}^{in}(\hat{x},b ,\mu)$ and ${\cal
C}^{out}(\hat{z}, b, \mu)$
contain contributions from partons radiated collinearily
to the initial and final hadrons.
These functions can also be expanded in series of
$\alpha_s/\pi$, as
\beq {\cal C}^{in}_{ij}(\xh, b, \mu) =
\sum_{k=0}^{\infty} {\cal C}^{in
(k)}_{ij}(\xh, \mu b)
\Biggl(\frac{\alpha_s (\mu)}{\pi} \Biggr)^k,
\eeq
\beq {\cal C}^{out}_{ij}(\zh, b, \mu) =
\sum_{k=0}^{\infty} {\cal C}^{out
(k)}_{ij}(\zh, \mu b) \Biggl(\frac{\alpha_s (\mu)}{\pi}\Biggr)^k.
\eeq
The renormalization scale in the ${\cal C}$-functions is
\bdm \mu= 2 e^{-\gamma}/b\equiv b_0/b, \edm where
$\gamma=0.577215...$ is the Euler constant.

The explicit expressions
for $A_k(C_1)$, $B_k (C_1, C_2)$ and the $\cal C$-functions can be
obtained by comparing the expansion of $W_{BA}(b, x, z, Q)$ as a series
in $\alpha_s/\pi$ with the
$b$-space expression for the perturbative cross-section. 
Using our NLO results, we find 
\bea
A_1 &=& C_F, \\
B_1 &=& 2 C_F \log \biggl(\frac{e^{-3/4} C_1}{b_0 C_2}\biggr).
\eea
To the same order, our expressions
for the  $\cal C$-functions are 
\begin{itemize}
\item LO:
\bea
{\cal C}^{in (0)}_{jk} (\xh, \mu b) &= &\delta_{jk} \delta (1-\xh),
\label{LO1}\\
     {\cal C}^{out (0)}_{jk} (\zh, \mu b) &= &\delta_{jk} \delta (1-\zh),\\
     {\cal C}^{in (0)}_{jg}=& {\cal C}^{out (0)}_{gj}&=0;
\eea
\item NLO:
\bea
{\cal C}^{in (1)}_{jk}(\hat{x},\mu b)&=& \frac{C_F}{2}(1-\xh) -
P_{qq}(\xh) \log \Bigl(\frac{\mu b} {b_0}\Bigr)  \nonumber\\
&-& C_F \delta(1-\xh) \biggl( \frac{23}{16}+ \log^2 \Bigl(
\frac{e^{-3/4} C_1}{ b_0 C_2} \Bigr) \biggr),\label{C1in}
\\
{\cal C}^{in (1)}_{jg}(\hat{x},\mu b)  &=& \frac{1}{2}\xh (1-\xh) -
P_{q G} (\xh) \log \Bigl(\frac{\mu b}{b_0}\Bigr),
\\
{\cal C}^{out (1)}_{jk}(\hat{z},\mu b) &=& \frac{C_F}{2}(1-\zh) -
P_{qq}(\zh) \log \Bigl(\frac{\mu b} {b_0}\Bigr) \nonumber \\
&-& C_F \delta(1-\zh) \biggl( \frac{23}{16}+ \log^2 \Bigl(
\frac{e^{-3/4} C_1}{ b_0 C_2} \Bigr) \biggr),\label{C1out}
\\
{\cal C}^{out (1)}_{gj}(\hat{z},\mu b) &=& \frac{C_F}{2}\zh -
P_{Gq} (\zh )\log \Bigl(\frac{\mu b}{b_0}\Bigr). \label{NLO4}
\eea
\end{itemize}
In these formulas, the indices $j$ and $k$ correspond to quarks and antiquarks,
and $g$ to gluons. Due to the crossing relations between
parton-level semi-inclusive DIS, vector boson production, and $e^+ e^-$
hadroproduction, the ${\cal C}^{in}$-functions are essentially
the same in semi-inclusive DIS and the Drell-Yan process;
and the
${\cal C}^{out}$-functions are essentially the same in semi-inclusive DIS and
$e^+ e^-$ hadroproduction.  At NLO the only difference stems from the
fact that the momentum transfer $q^2$ is spacelike in DIS and timelike
in the other two
processes. The virtual diagrams Figs.~\ref{diags}b-d differ by
 $\pi^2$ for spacelike and timelike $q^2$. Correspondingly, ${\cal
C}^{in (1)}_{jk}$ and ${\cal C}^{out (1)}_{jk}$
for semi-inclusive DIS
(\ref{C1in}) do not
contain the term $(\pi^2/3) \delta (1-\xh)$ (or $(\pi^2/3) \delta
(1-\zh)$ ) which is present in
the ${\cal C}^{in (1)}_{jk}$-function for the Drell-Yan process
(or in the ${\cal
C}^{out (1)}_{jk}$-function for $e^+ e^-$
hadroproduction, respectively)\footnote{With two minor exceptions,
our expressions for the functions ${\cal C}$ are
equivalent to the ones published previously by {\it Meng et al.} 
\cite{Meng2}.
The
 $\pi^2/3$ terms are incorrectly included
in equations (43) and (45) of \cite{Meng2} for the
semi-inclusive DIS functions ${\cal
C}^{in (1)}_{jk}$ and  ${\cal C}^{out (1)}_{jk}$.
Also, our Eq.  (\ref{C1out}) contains $23 C_F/16 = 23/12$ instead
 of 29/12 in Eq.  (45) of \cite{Meng2}, which is apparently due to a typo
in \cite{Meng2}.}.  On the other hand, the NLO expression for the Sudakov
factor (\ref{sud_p}) is the same for semi-inclusive DIS, the Drell-Yan
process, and $e^+ e^-$ hadroproduction, which also results from the
crossing symmetry.

Up to now, we have been discussing the behavior of the resummed
cross-section at short distances.
The representation (\ref{W}) should be modified at large values of the
the variable $b$ to account for  nonperturbative long-distance
dynamics. The modified ansatz for $W_{BA}$ valid at all values of $b$ is
\begin{equation}\label{W2}
W_{BA}(b, x, z, Q)=\sum_j e_j^2
(D_{B/b} \circ {\cal C}^{out}_{bj})(z, b_*, \mu) ({\cal
C}_{ja}^{in} \circ F_{a/A}) (x, b_*, \mu)~e^{-S_{BA}}.
\end{equation}
Here the variable
\beq
 b_* \equiv \frac{b}{\sqrt{1 + \Bigl( \frac{b}{b_{max}}\Bigr)^2}}
\label{bstar}
\eeq
serves to turn off the perturbative dynamics  for $b
\geq b_{max}$, with $b_{max}~\approx~ 1~\mbox{GeV}^{-1}$.
Furthermore, the Sudakov factor is modified, being written as the sum of
the perturbatively calculable part $S^P (b_*, Q)$
given by (\ref{sud_p}), and a
 nonperturbative part which is only partially constrained by the theory:
\beq
 S_{BA} (b, x, z, Q) = S^P (b_*, x, z, Q) + S^{NP}_{BA} (b, x, z, Q).
\label{SPNP}
\eeq
From the renormalization properties of the theory, it can be concluded that
the $Q$ dependence of the nonperturbative Sudakov term should be separated
from the dependence on the other kinematic variables, {\it i.e.}
\beq
S^{NP}_{BA} (b, x, z, Q) =
g_{BA}^{(1)} (b, x, z) +  g_{BA}^{(2)} (b) \log \frac{Q}{Q_0},
\eeq
with $Q_0 \approx 1 \ \mbox{GeV}$. The theory does not predict the functional
forms of $g_{BA}^{(1)} (b, x, z)$ and $g_{BA}^{(2)} (b)$, so these must be
determined by fitting  experimental data.
Also, $S^{NP}_{BA}$ can depend
on the types of the hadrons $A$ and $B$.  On the other hand, due to the
crossing symmetry between semi-inclusive DIS, the Drell-Yan process and
$e^+ e^-$ hadroproduction, one may expect that the functions $g^{(2)} (b)$
in these processes are not independent~\cite{Meng2}, but satisfy
  \beq \label{g2}
\left. g_{BA}^{(2)}(b) \right|_{sDIS} =
\frac{1}{2}\Bigl(\left. g_{AA}^{(2)}(b) \right|_{DY} +
\left. g_{BB}^{(2)}(b) \right|_{e^+ e^-}\Bigr).
\eeq
If the  relationship (\ref{g2}) is true, then the function
$g_{BA}^{(2)} (b)$ in
semi-inclusive DIS is completely known once the parameterizations for
the
functions
$g_{AA}^{(2)} (b)$ in the Drell-Yan
and $g_{BB}^{(2)} (b)$ in $e^+ e^-$ hadroproduction processes are
available.
In practice, the Drell-Yan nonperturbative Sudakov factor
is known only when the incoming particle is a nucleon
\cite{DWS,LY,BY,BrY}, while the
nonperturbative Sudakov factor for  $e^+ e^-$ hadroproduction is
available only for  energy correlations \cite{CS}.
An additional complication comes from the fact that the known
parameterizations of the nonperturbative Sudakov factors for the
Drell-Yan \cite{DWS,LY,BY,BrY}
and $e^+e^-$ hadroproduction \cite{CS} processes
correspond to  slightly different scale choices,
\beq
C_1= b_0,\quad C_2=1
\label{C121}
\eeq
and
\beq
C_1= b_0,\quad C_2=e^{-3/4},
\label{C122}
\eeq
respectively.
Therefore, the known functions $\left. g^{(2)}\right|_{DY} (b)$ and
$\left. g^{(2)}\right|_{e^+e^-} (b)$ are not 100\% compatible, and in
principle should not be combined to obtain $g^{(2)}(b)$ for
semi-inclusive DIS.

Despite this minor incompatibility,
we will use (\ref{g2}) to construct $g^{(2)}(b)$ for our
numerical calculation of  energy flows and, with less justification,
particle multiplicities.  We have found that the numerical results for the
energy flows are only slightly dependent on the choice between the
two sets (\ref{C121},\ref{C122}) of the constants $C_1,\ C_2$
(see Section~\ref{6}). Also,  detailed information about  the functional
form of the $Q$-dependent part of the nonperturbative Sudakov factor
can be obtained only by studying the dependence of the resummed
cross-sections on $Q$. Since the HERA data discussed in this paper
covers only a small range of $Q$ ($3.62 \leq Q \leq 5.71 \ \mbox{GeV}$), it is
hard to distinguish between the uncertainties in the $Q$-dependent and
constant parts of
the nonperturbative Sudakov factor.
Of course, more definite conclusions about $g^{(2)}$ will
be possible once more detailed semi-inclusive DIS
data from HERA and Fermilab-E665, covering a wider range of $Q$,
are available.

\section
{Hadronic multiplicities and energy flows \label{4}}

Knowing the hadron cross-section, it is possible to calculate the multiplicity
of the process, which is defined as the ratio of this cross-section, and the
total inclusive DIS cross-section for the given leptonic cuts:
\beq
\mbox{Multiplicity} =
\frac{1}{d \sigma_{tot}/d x d Q^2}\frac{d \sigma}{d x d z
d Q^2 d q_T^2}.
\eeq
Both the cross-section and the multiplicity depend on the properties of the
final-state fragmentation. The analysis can be simplified by
considering energy flows which do not have such dependence. A traditional
variable used in the experimental literature is a transverse energy
flow
$\langle E_T \rangle$ in one of the coordinate frames, defined as
\beq  \label{ET} \langle E_T \rangle_{\Phi_B} = \frac{1}{
\sigma_{tot}}\sum _B \int_{\Phi_B} d \Phi_B \  E_T \frac{ d\sigma (e +A \ra
e+ B+ X)} {d \Phi_B}.  \eeq
This definition involves an integration over
the available phase space $\Phi_B$ and a summation over all possible
species of the final hadrons $B$. Since the integration over $\Phi_B$
includes integration over the longitudinal component of the
momentum of $B$, the dependence of $\langle E_T \rangle$ on the fragmentation functions
drops out due to the normalization condition
\beq
\sum_B \int\ z\ D_{B/b} (z) d z = 1.
\eeq

Instead of $\langle E_T \rangle$, we analyze
the flow of the variable $z$. This flow is  defined as
\beq \frac{d
\Sigma_z}{d x \ dQ^2 \ dq_T^2} = \sum_B \int_{z_{min}}^1 \ z \ \frac{
d\sigma(e +A \ra e+ B+ X) }{dx \ dz \ dQ^2 \ dq_T^2}\ d z. \eeq
We prefer to use $\Sigma_z$ rather than $\langle E_T \rangle$
because $\langle E_T \rangle$
is not Lorentz invariant, which complicates its
usage in the theoretical analysis\footnote{The $z$-flow $\Sigma_z$
is related to
the energy distribution function $\Sigma$ calculated in \cite{Meng2} as
$\Sigma_z =(2 x E_A/Q^2)\Sigma$.
Here $E_A$ is the energy of the initial hadron in the HERA lab frame.}.
Also, the analysis in terms of
$q_T$ and $z$-flow makes the analogy between resummation in the
current region of  semi-inclusive DIS and in the small transverse
momentum region of the Drell-Yan process more obvious.
 Since $q_T$ is simply related to the
pseudorapidity in the $\vg p$ \cms   frame via Eq. (\ref{qTetacm}), and
the transverse energy of a nearly massless particle in this frame is
given by
\beq
  E_T \approx p_T = z q_T,
\eeq
the experimental information on $d\Sigma_z/ (d x \ dQ^2 \ dq_T^2)$
can be derived from the $\gamma^* p$ \cms    frame
pseudorapidity ($\eta_{cm}$) distributions of
$\langle E_T \rangle$ in bins of $x$ and $Q^2$. If
mass effects are neglected, we have
\beq
\frac{d \langle E_T \rangle}{d\eta_{cm}} = q_T^2 \frac{d \Sigma_z}{d q_T}.
\label{Sz2ET}
\eeq

The asymptotic contribution to the
  $z$-flow distribution  is
\bea
\Biggl(\frac{d \Sigma_z}{ d x d Q^2 d q_T^2 d\phi}
\Biggr)_{asym}&=& \sFs \alpi \frac{e^2}{2 q_T^2}  \frac{A_1(\psi, \phi)}{2
\pi}
\nonumber \\
&\times& \sum_{j}
e_j^2 \Biggl[ \Bigl\{ (P_{qq} \circ f_{j/A})(x,\mu)
+(P_{qg}\circ f_{g/A})(x,\mu)\Bigr\}
\nonumber \\
&+& 2 f_{j/A}(x,\mu)  \Bigl\{C_F \log
\frac{Q^2}{q_T^2}-\frac{3}{2}C_F\Bigr\} +{\cal
  O}\Bigl(\biggl(\frac{\alpha_s}{\pi}\biggr)^2, q_T^2\Bigr)\Biggr].
\label{asymz}
\eea
The resummed $z$-flow distribution is
\begin{equation}
\label{resumz}
\frac{d \Sigma_z}{d x d Q^2 d q_T^2 d \phi}=
\sFs \frac{ A_1 (\psi, \phi)}{2}
\int \frac{d^{n-2}b}{(2\pi)^{n-2}} e^{i\vec q_T \cdot \vec
b } W_z (b, x, Q)+Y_z,
\end{equation}
with
\begin{equation}\label{Wz}
W_z (b, x, Q)=\sum_j e_j^2
{\cal C}^{out}_z (b_*, \mu) ({\cal
C}_{ja}^{in} \circ F_{a/A}) (x, b_*, \mu)~e^{-S_z (b, x , Q)}.
\end{equation}
In (\ref{Wz}), the NLO function $ {\cal C}^{out}_z (b, \mu) $ is
\bdm
 {\cal C}^{out}_z (b, \mu) = \Biggl[ 1 +
\frac{\alpha_s}{\pi} C_F \Bigl( -\frac{65}{48} +
\frac{4}{3}\log \frac{b\mu}{b_0} -
\log^2 \frac{e^{-3/4} C_1}{C_2 b_0}\Bigr)\Biggr] +
\frac{\alpha_s}{\pi} \frac{C_F}{2} \Biggl[ \frac{1}{3} -
\frac{8}{3}\log \frac{b \mu}{b_0}\Biggr].
\edm

Similar to (\ref{SPNP}), the  $z$-flow Sudakov factor $S_z$ is a sum
of  perturbative and nonperturbative parts,
 \beq
 S_{z} (b, x, Q) = S^P (b_*, x, Q) + S^{NP}_{z} (b, x, Q).
\label{SzPNP}
\eeq
The NLO perturbative Sudakov factor $S^P$ is given by the universal
$x$-independent expression (\ref{sud_p}). By the same argument
as in the case of semi-inclusive DIS multiplicities, we assume that the
non-perturbative part of $S_z$ can be parameterized as
 \beq
S^{NP}_{z} (b, x, Q) =
g^{(1)} (b, x) +
\frac{1}{2}\Bigl(\left. g^{(2)}(b) \right|_{DY} +
\left. g^{(2)}(b) \right|_{e^+ e^-}\Bigr)
 \log \frac{Q}{Q_0}.
\label{SzNP}
\eeq
In the numerical  calculation, we use the functions
$\left. g^{(2)}(b)\right|_{DY}$ from \cite{DWS} and
$\left. g^{(2)}(b)\right|_{e^+ e^-}$ from \cite{CS}, despite the fact
that  $\left. g^{(2)}(b)\right|_{DY}$ was fitted to Drell-Yan data
using a different $C_2$
value than  $\left. g^{(2)}(b)\right|_{e^+ e^-}$.

We also parameterize
the functional form of $g^{(1)} (b, x)$ in terms of $b$ and $x$ as
\beq
g^{(1)} (b, x) = ( h_{1} + \frac{h_{2}}{\sqrt{x}} ) b^2,
\eeq
where the constants $h_1,\ h_2$ must be determined
by fitting the experimental data.

In principle, the $z$-flow Sudakov factor $S_z (b, x , Q)$
is related to  the Sudakov factors $S_{BA}
(b, x, z, Q)$ of the  contributing hadroproduction
processes $ e +A \ra e + B + X$ through the relationship
\beq
e^{-S_z (b, x)} = \frac{1}{C_z^{out}(b_*, \mu)}
\sum_B \int z dz e^{-S_{BA} (b, x, z, Q)}
(D_{B/b} \circ C_{b,j}^{out})(z, b_*, \mu).
\label{SS}
\eeq
In practice, the efficient usage of this relationship to constrain
the Sudakov factors is only
possible if the fragmentation functions and the hadronic contents of
the final state are accurately known. We do not use the relationship
(\ref{SS}) in our calculations.

\section{Relationship between the
perturbative and resummed cross-sections. Uncertainties of the
calculation \label{5}
}

In the numerical calculations, some
care is needed to treat the uncertainties in the definitions of the
asymptotic and resummed cross-sections (\ref{asym}) and
(\ref{resum}), although formally these uncertainties are of order
${\cal O}((\alpha_s /\pi)^2, q_T^{-1})$.

\subsection{Matching}
\begin{figure}
\epsfxsize 14cm
\epsffile{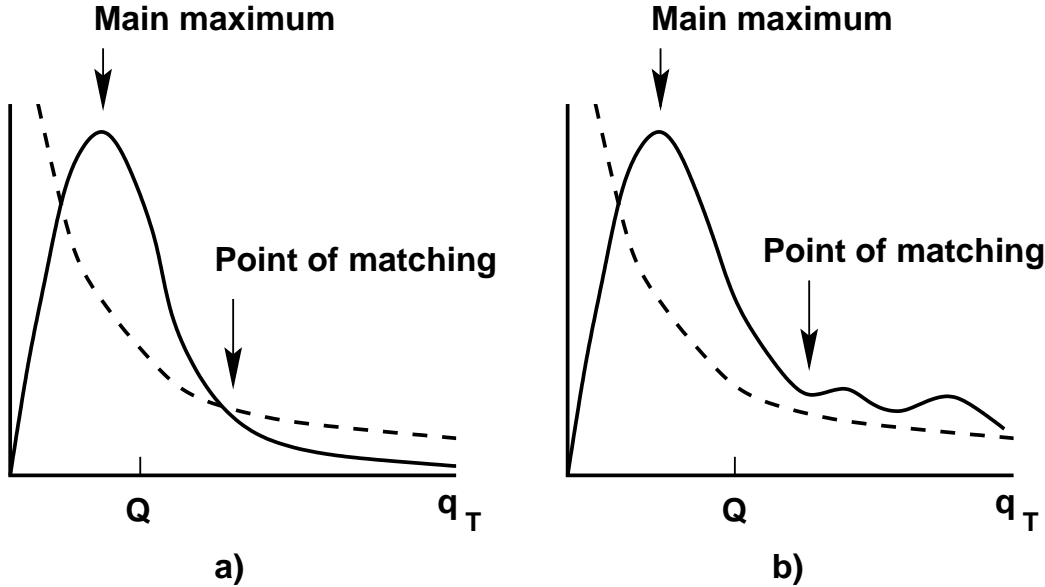}
\caption{\label{matching}
Choice of the matching between the resummed and perturbative
cross-section}
\end{figure}

The generic structure of the resummed cross-section (\ref{resum}),
calculated up to the order ${\cal O} ((\alpha_s /\pi)^N)$, is
\beq
\sigma_{resum}^{(N)} = \sigma_{CSS} + Y^{(N)}.
\label{r}
\eeq
In (\ref{r}), the CSS piece receives all-order contributions from large
logarithmic terms
\beq
\frac{\alpha_s^n}{q_T^2} \sum_{m=0}^{2 n -1}
v_{mn}  \log^m{(q_T^2/Q^2)}, \quad n=1,..., \infty.
\eeq
The $Y$-piece is
the difference of the fixed-order perturbative and asymptotic
cross-sections,
\beq
Y^{(N)} = \sigma_{pert}^{(N)} - \sigma_{asym}^{(N)}.
\label{Y2}
\eeq

In the small-$q_T$ region, we expect cancellation up to terms of
order
${\cal O}(\alpha_s^{N+1}/\pi^{N+1})$ between the
perturbative and asymptotic pieces in (\ref{Y2}), so that the CSS
piece dominates the resummed cross-section (\ref{r}). On the other
hand, the expression for the asymptotic piece coincides with the
expansion of the CSS piece up to the order ${\cal
O}(\alpha_s^N/\pi^N)$.
Therefore, at larger $q_T$,
we expect better cancellation between the CSS and
asymptotic pieces, and at large $q_T$ the resummed cross-section
(\ref{resum})
should be equal to the perturbative cross-section up to corrections of
order ${\cal O}(\alpha_s^{N+1}/\pi^{N+1})$.

In principle, due to the cancellation between the perturbative and
asymptotic pieces at small $q_T$, and between the resummed and
asymptotic piece at large $q_T$, the resummed formula
$\sigma_{resum}$ is at least as  good an approximation of the physical
cross-section as the perturbative cross-section $\sigma_{pert}$
of the same order.
However, in the NLO calculation at large $q_T$
it is safer  to use the fixed order cross-section (\ref{hadcs})
instead the resummed expression (\ref{resum}).
At the NLO order of $\alpha_s$, the difference between the CSS
and the asymptotic pieces at large $q_T$ may still be
non-negligible.
Therefore, the resummed cross-section $\sigma_{resum}$
may differ significantly from the NLO perturbative
cross-section $\sigma_{pert}$.
This difference does not mean that the
resummed cross-section agrees with the data better than the
fixed-order one. At $q_T \approx Q$, the NLO cross-section is no
longer dominated by the logarithms that are resummed in
(\ref{resum}). In other words, the resummed cross-section (\ref{resum})
does not include some terms in the NLO cross-section that become
important at $q_T \approx Q$.
For this reason, at large $q_T$ the resummed cross-section may show unphysical
behavior; for example, it can become negative or oscillate.

As the order of the
perturbative calculation increases, we expect the agreement between
the resummed  and the fixed-order perturbative
cross-sections to improve.
Indeed, such improvement was shown in the case of vector boson
production \cite{BY}, where one observes a smoother transition from the
resummed to the fixed-order perturbative cross-section if the
calculation is done at the next-to-next-to-leading order.
Also, at the NNLO the switching occurs at
larger values of the transverse momentum of the vector boson than in the case
of the NLO.

The switching from the resummed to the fixed-order
perturbative cross-section should occur at
$q_T \approx Q$. Nonetheless,
there is no unique prescription for the exact point at which it
should happen.
In our program we switch from the resummed cross-section
to the perturbative result
at the first
minimum of the difference
\bdm \Biggl(\frac{d \sigma}{d q_T}\Biggr)_{resum} -
\Biggl(\frac{d \sigma}{d q_T}\Biggr)_{pert},
\edm
lying  at $q_T$ above the main maximum of the resummed cross-section
(Fig.~\ref{matching}).   This prescription is satisfactory for two
different possible situations, in which the resummed cross-section
crosses (Fig.~\ref{matching}a) or does not cross (Fig.~\ref{matching}b)
the perturbative cross-section.

\subsection{Kinematic corrections at $q_T \approx Q$}
In this subsection we will discuss the differences between
the kinematics implemented in the definitions of the asymptotic and
resummed cross-sections (\ref{asym}) and (\ref{resum}), and the
kinematics of the perturbative piece at non-zero values of $q_T$.

\begin{figure}[H]
\epsfxsize 18cm
\epsffile {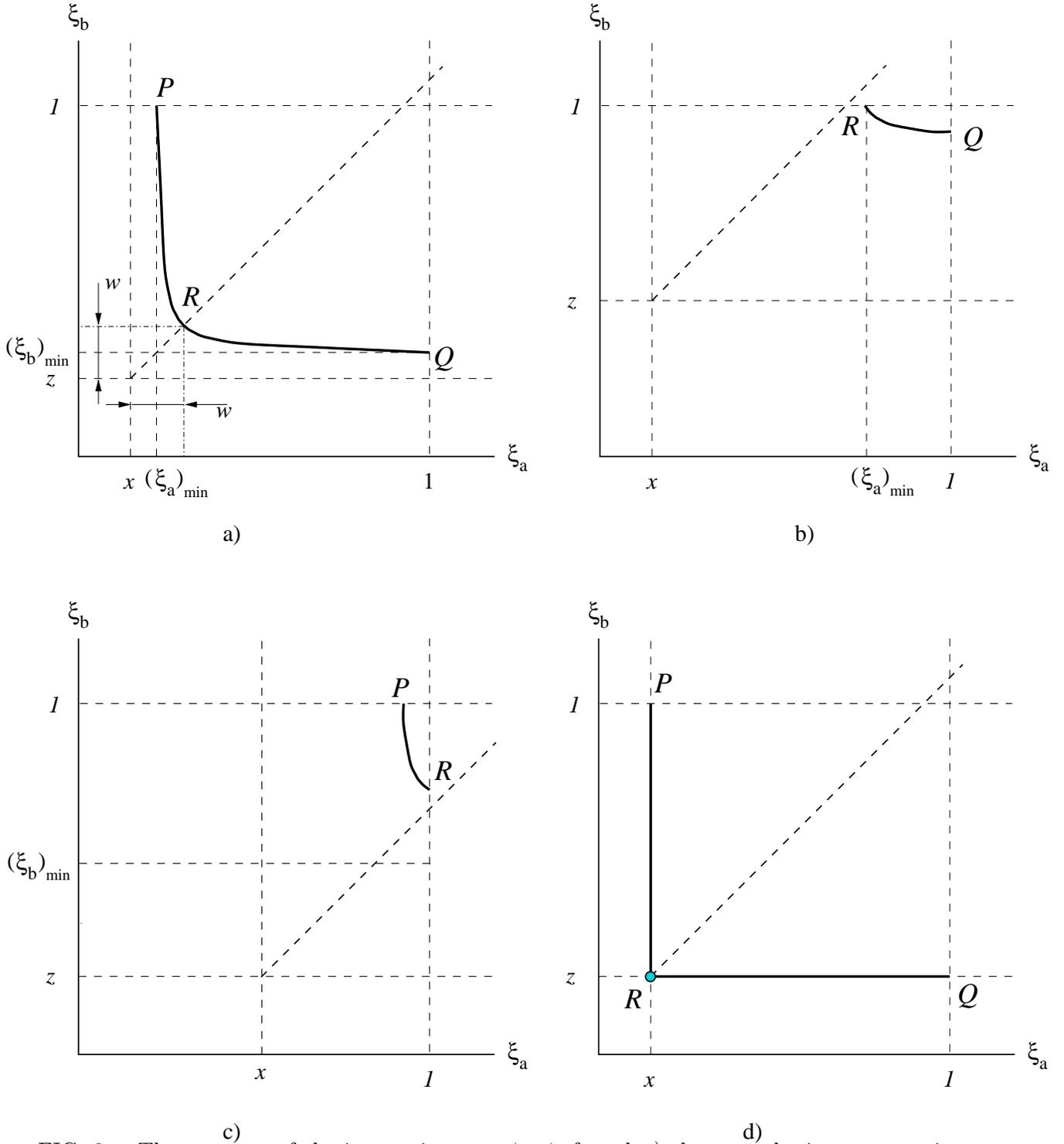}
\caption{\label{xiaxib} The contours of the integration over $\xi_a$,
$\xi_b$ for a,b,c) the perturbative cross-section (\ref{hadcs}); d) the
asymptotic and resummed cross-sections (\ref{asym}) and (\ref{resum}).
}
\end{figure}

Consider the perturbative hadronic cross-section (\ref{hadcs}). The
integrand of (\ref{hadcs}) contains the delta-function
\beq
\label{cond}
\delta \Biggl[ \frac{q_T^2}{Q^2} - \biggl(\frac{1}{\xh}-1\biggr)
\biggl(\frac{1}{\zh}-1\biggr)\Biggr]=
x z \delta\Biggl[(\xi_a- x ) (\xi_b -z ) - x z \frac{q_T^2}{Q^2}\Biggr]
\eeq
which comes from the parton-level cross-section (\ref{sighat}).
Depending on the values of $x, z, Q^2, q_T^2$, the contour of
the integration over $\xi_a$ and $\xi_b$ determined by (\ref{cond})
can have one of three shapes shown in  Fig.~\ref{xiaxib}a,b,c.
For $q_T \ll Q$ the integration proceeds along the contour in
Fig.~\ref{xiaxib}a, and the integral in (\ref{hadcs}) can be written
in either of two alternative forms
\bea
\frac{d \sigma_{BA}}{d x d z d Q^2 d q_T^2 d\phi} &=&
\int_{(\xi_a)_{min}}^{1} \frac{d \xi_a}{\xi_a - x}
M_{BA}(\xi_a, \xi_b; \xh, \zh, Q^2, q_T^2, \phi) \nonumber \\
                                          & & \nonumber \\
&=& \int_{(\xi_b)_{min}}^{1} \frac{d \xi_b}{\xi_b - z}
M_{BA}(\xi_a, \xi_b; \xh, \zh, Q^2, q_T^2, \phi),
\eea
where
\bdm
M_{BA}(\xi_a, \xi_b; \xh, \zh, Q^2, q_T^2, \phi) =
\edm
\beq
\frac{ \sigma_0 F_l}{4 \pi \SeA Q^2} \frac{\alpha_s}{\pi}\xh \zh
\sum_{a,b,j}
e^2_j D_{B/b}(\xi_{b}) F_{a/A}(\xi_{a})
\sum_{\alpha=1}^{4} f^{(\alpha)}_{ba}(\xh, \zh, Q^2, q_T^2)
A_{\alpha}(\psi, \phi).
\eeq
The lower bounds of the integrals are
\beq
(\xi_a)_{min}= \frac{w^2}{1 - z} +x,
\eeq
\beq
(\xi_b)_{min}= \frac{w^2}{1 - x} +z,
\eeq
with
\beq
 w \equiv \frac{q_T}{Q} \sqrt{x z}.
\eeq

Alternatively, the cross-section can be written in a form symmetric with
respect to $x$ and $z$,
\bea
\frac{d \sigma_{BA}}{d x d z d Q^2 d q_T^2 d\phi} &=&
\int_{x + w}^{1} \frac{d \xi_a}{\xi_a - x}
M(\xi_a, \xi_b; \xh, \zh, Q^2, q_T^2,\phi)  \nonumber \\
 & &  \nonumber \\
 &+& \int_{z + w}^{1} \frac{d \xi_b}{\xi_b - z}
M(\xi_a, \xi_b; \xh, \zh, Q^2, q_T^2,\phi),
\label{sym}
\eea
where the integrals are calculated along the branches $RP$ and $RQ$ in
Fig.~\ref{xiaxib}a, respectively.
As $q_T \ra 0$,
\bdm
(\xi_a)_{min} \ra x,\ (\xi_b)_{min} \ra z,
\edm
and the contour $PRQ$ approaches the contour of integration of the asymptotic
cross-section (\ref{asym}) shown in Fig.~\ref{xiaxib}d.  The
horizontal  (or vertical)
branch contributes to the convolutions with splitting
functions in (\ref{asym})
arising from the initial (or final) state collinear
singularities, while the soft singularities of (\ref{asym}) are located at
the point $\xi_a = x,\ \xi_b = z$.

On the other hand, as $q_T$ increases up to values around $Q$, the
difference between the contours of integration of the perturbative and
asymptotic cross-sections may become significant.
First, as can be seen from (\ref{sym}), in the perturbative piece $\xi_a$ and
$\xi_b$ are always higher than $x+ w$ or
 $z + w$ , while in the asymptotic piece they
vary between $x$ or $z$ and~unity. At small $x$ (or small $z$) the
difference between the phase spaces of the perturbative and asymptotic pieces
may become important due to the steep rise of the PDFs and FFs in this
region. Indeed, for illustration
consider a semi-inclusive DIS experiment at small $x$.
Let $q_T/Q = 0.5$, $z=0.1$, and  $x=10^{-4}$; then
$x+ w = 1.6 \cdot 10^{-3} \gg x = 10^{-4}$.
In combination with the fast rise of the PDFs at small $x$,
this will enhance the
difference between the perturbative and asymptotic cross-sections.

Second, for $x$ or $z$ near unity, it could happen that
$x + w \geq 1$ or $ z+ w \geq 1$, which
would lead to the disappearance of one or two branches of the
integration of the
 perturbative piece (Fig.~\ref{xiaxib}b,c). In this
situation the phase space for nearly collinear radiation along the
direction of the initial or final parton is suppressed. Again, this may
degrade the consistency between the perturbative and asymptotic piece, since
the latter includes contributions from both branches of the collinear
radiation.
Fortunately, the $x-z$ asymmetry of the phase space in
semi-inclusive DIS is not important in the analysis of the existing data
from HERA, since it covers the small-$x$ region and is less sensitive to
the contributions from the large $z$ region, where the rate of the
hadroproduction is small. However, in the numerical  analysis we found it
necessary to correct for the contraction of the perturbative phase space
described in the previous paragraph. We incorporate this correction by
substituting for $x$ and $z$ in (\ref{asym}, \ref{resum}) the rescaled
variables
\bdm
\tilde x = \frac{Q^2 +q_T^2}{Q^2} x,
\edm
\beq
\tilde z =
\frac{Q^2 +q_T^2}{Q^2} z.  \label{tildexz}
\eeq
These substitutions
simulate the phase space  contraction of the perturbative piece. At small
$q_T$, the rescaling
reproduces the exact asymptotic and resummed pieces (\ref{asym})
and (\ref{resum}), but at larger $q_T$ it excludes the
unphysical integration regions of $\xi_a \approx x $ and $\xi_b \approx z$.
\newpage
\newpage

\section{Numerical results \label{6}}
In this section, we present the results of Monte-Carlo simulations for
the $z$-flow and the differential multiplicity of charged particle
production,
 \bdm \frac{1}{d\sigma_{tot}/(dx dQ^2)} \frac{d \Sigma_z}{dx dQ^2 d
q_T} \mbox{ and } \frac{1}{d\sigma_{tot}/(dx dQ^2)} 
\frac{d \sigma^{chgd}}{dx dz
dQ^2 d q_T}.  \edm
Our calculations use the parameters of the HERA
electron-proton collider.  The energies of the proton and electron beams
are taken to be equal to 820 and 27.5 \mbox{GeV}, respectively.

\subsection{Energy flows}
As a first application of the resummation formalism, we consider the \cms
pseudorapidity distributions of the transverse energy flows in the current
region, data for which has been  published in \cite{H1z}. We discuss the
data in seven bins of $x$ and $Q$, four of them covering the region $10
\leq Q^2 \leq 20\ \mbox{GeV}^2$, $3.7 \cdot 10^{-4} \leq x \leq 2.3 \cdot
10^{-3}$, and the other three the region $20 \leq Q^2 \leq 50 \
\mbox{GeV}^2$, $9.3 \cdot 10^{-4} \leq x \leq 4.9 \cdot 10^{-3}$.

\begin{figure}[h]
\epsfysize 10cm
\epsffile{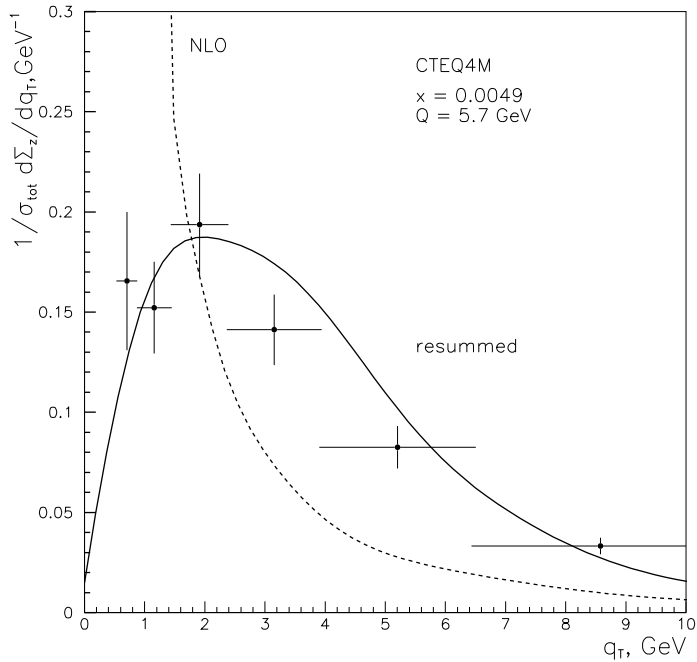}
\caption{\label{nloandresum}
Comparison of the NLO perturbative (\ref{zx})
and resummed (\ref{resumz})
expressions for the $z$-flow distribution with the existing
experimental data from HERA \protect\cite{H1z}.
The presented data is for
$\langle x\rangle =0.0049, \ \langle Q^2 \rangle=32.6 \ \mbox{GeV}^2$.
}
\end{figure}
\newpage
\begin{figure}[H]
\epsfysize 18cm
\epsffile{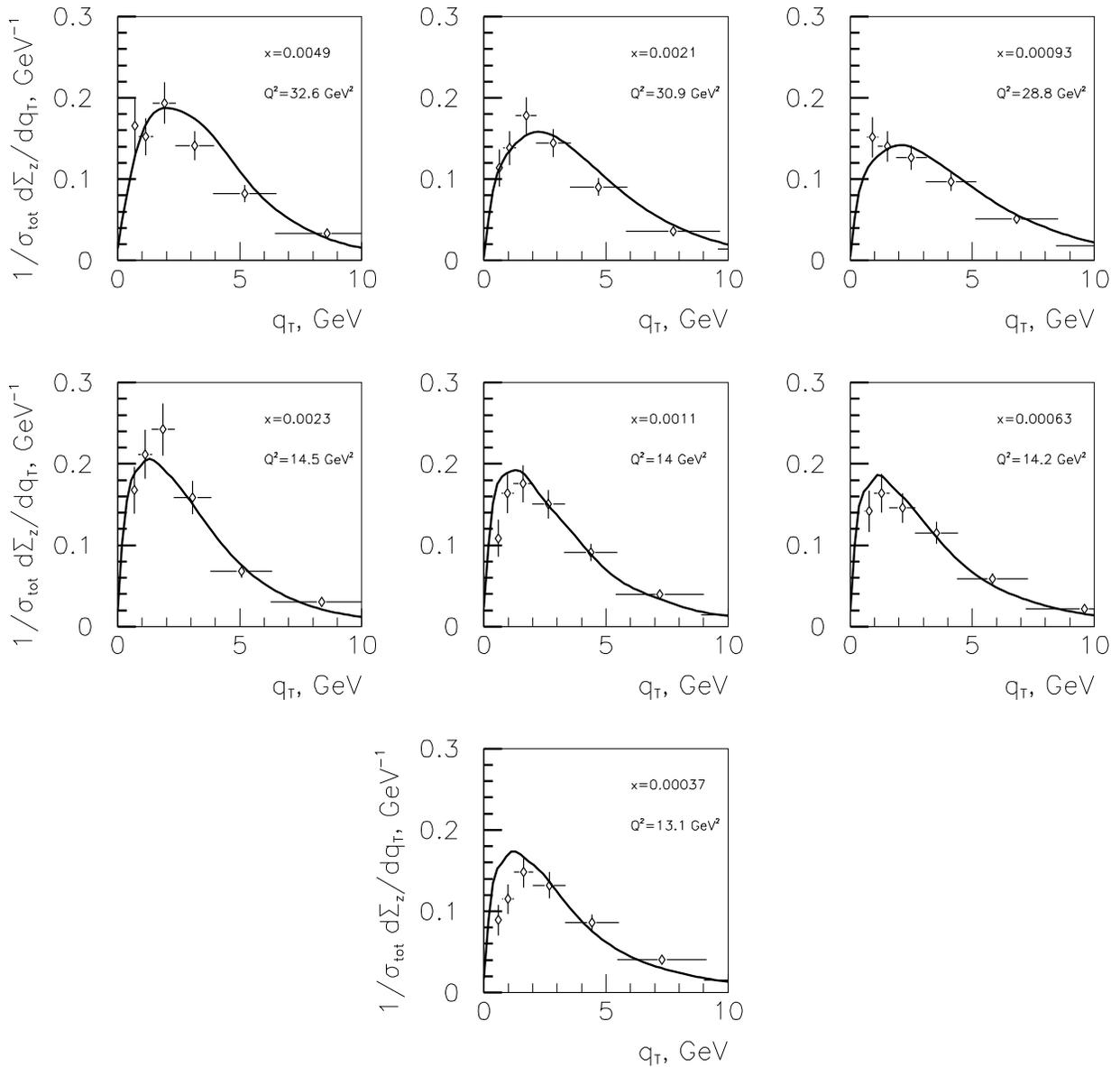}
\caption{ \label{zqt}
Comparison  of the  resummed (\ref{resumz})
$z$-flow distribution with the HERA data from {\protect\cite{H1z}}
in 7 bins of $\langle x \rangle$ and $\langle Q\rangle $. The resummed
$z$-flows were calculated using the parameterization (\ref{SzNP2}) for the
non-perturbative part of the Sudakov factor.  } \end{figure}

In our calculation, we use the CTEQ4M parton distribution functions
\cite{CTEQ}. The factorization and
renormalization scales of the perturbative and asymptotic pieces
are all set equal to $\mu = Q$.
As was mentioned in Section \ref{4}, the data on transverse energy flow
can be easily  transformed into the $q_T$ distributions
of the $z$-flow. In
Fig.~\ref{nloandresum},
we present the comparison of  the existing data in
one of the bins of $x$ and $Q^2$
($\langle x\rangle =0.0049, \ \langle Q^2 \rangle = 32.6 \ \mbox{GeV}^2$)
with the NLO perturbative
and resummed $z$-flows given in Eqs. (\ref{zx}) and (\ref{resumz}),
respectively.
In Fig.~\ref{zqt} we present the comparison of the resummed $z$-flow
with the data in the other bins of \cite{H1z}.

Figure~\ref{nloandresum}
demonstrates two important aspects of the NLO $q_T$ distribution,
namely, that the NLO exceeds the data at small $q_T$ and is below the
data at $q_T \geq Q$. In fact, we find that the deficit of the
NLO prediction of  perturbative theory in comparison with the data
at medium and large $q_T$ ($q_T \geq 5 \ \mbox{GeV}$)
is present in the entire region of  $x$ and
$Q^2$ that we have studied.

As we discussed in Section \ref{5}, one can trust
the resummed calculation only for reasonably small values of
$q_T/Q$. For large values of $q_T$, the fixed-order perturbative
result is more reliable. This means that
the NLO resummation formalism will not give an accurate description of
the data for $q_T \gg Q$, due to the small magnitude of the NLO
perturbative $z$-flow in this region.

The excess of the data over the NLO calculation can be interpreted
as a  signature of
other intensive hadroproduction mechanisms
at \cms   pseudorapidities $\eta^{cm} \leq 2$. A discussion of
the cross-sections in this pseudorapidity region is beyond
the scope of our paper. However, we would like to point out that there
exist several possible explanations of the data in this region,
for instance,
the enhancement of the cross-section due to  BFKL showering
\cite{BFKL} or  resolved photon contributions \cite{Kramer,Jung}.
From the point of view of our study, it is clear that
good agreement between the data and the combination of the
perturbative calculation and the CSS resummation, in a wider range of
$\eta^{cm}$, will be achieved when
next-to-next-to-leading order contributions, like the ones
contributing to  (2+1) jet production \cite{Catani},
are taken into account.

On the other hand, Figs.~\ref{nloandresum} and \ref{zqt} illustrate that
the resummation formalism  accurately describes
the data in the region $q_T \leq 10 \ \mbox{GeV}$. This calculation of the
resummed $z$-flows (\ref{resumz}) was done using the following
parameterization for the non-perturbative Sudakov factor (\ref{SzNP})
\beq
S^{NP}_{z} (b, x, Q) =  g^{(1)}(b,x) +
g^{(2)}(b, Q),
\label{SzNP2}
\eeq
where the $Q$-dependent part $g^{(2)}(b, Q)$
is completely defined by the symmetry between  semi-inclusive  DIS,
Drell-Yan and $e^+ e^-$ hadroproduction processes (Section \ref{3}),
\beq
g^{(2)}(b,Q) =
\frac{1}{2}b^2\Biggl( 0.48 \log (\frac{Q}{2 Q_0}) +
5.32 C_F
\log\Bigl(\frac{b}{b_*}\Bigr) \log\Bigl(\frac{C_2 Q}{C_1 Q_0}\Bigr)
\Biggr).
\label{g2_2}
\eeq
The parameterizations of the $Q$-dependent parts of the
non-perturbative Sudakov factors in the Drell-Yan and $e^+ e^-$
hadroproduction processes are taken from \cite{DWS} and \cite{CS},
respectively.
In (\ref{g2_2}), the constants are
$C_1 = 2 e^{-\gamma},\ C_2 = e^{-3/4}, Q_0 = 1 \ \mbox{GeV}$. The variable
$b_*$ is given by (\ref{bstar}), with $b_{max} = 1 \ \mbox{GeV}^{-1}$.

\begin{figure}
\epsfysize 10cm
\epsffile{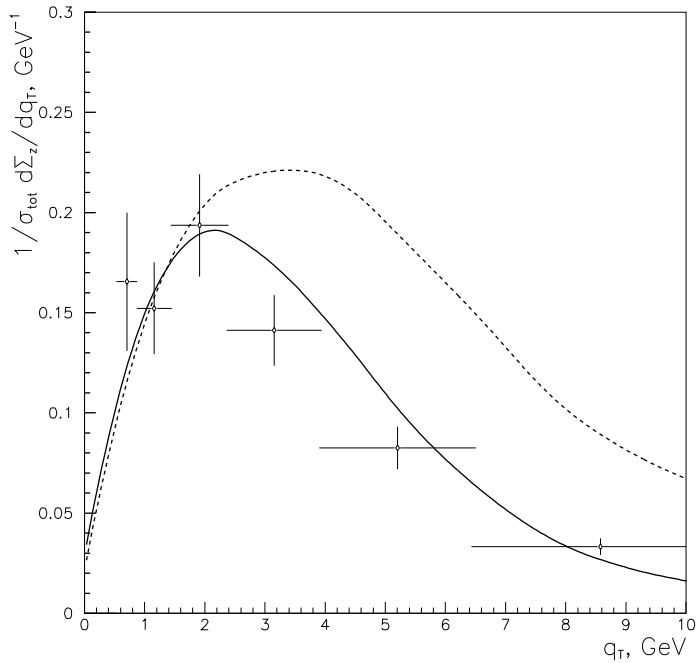}
\caption{\label{ixqtfl_dep} The comparison of the $z$-flow distributions for
$\langle x \rangle =0.0049 \mbox{ and } 
\langle Q^2 \rangle = 32.6 \ \mbox{GeV}^2$,
calculated with (solid) and without (dashed)
the kinematic correction (\ref{tildexz}).
}
\end{figure}
The $x$-dependent function $g^{(1)}(x,Q)$ was
determined by fitting the HERA data of \cite{H1z}. We found that good
agreement with the data is obtained when $g^{(1)}(b,x)$ is parameterized by
a linear function of $x^{-0.5}$,
\beq
g^{(1)}(b,x) = b^2 \biggl(-2.58 +
\frac{0.58}{\sqrt{x}}\biggr), \quad \mbox{ for } x \leq 10^{-2}.
\label{g1_2}
\eeq
We should emphasize that we don't know the exact
 behavior of $g^{(2)}(b,x)$ for $x \geq 10^{-2}$, where currently there is
no data available.  On the other hand, the term $g^{(2)}(b,x)$ in the
parameterization (\ref{g1_2}) becomes negative for $ x \geq 5 \cdot
10^{-2}$ which makes the numerical value of the resummed $z$-flow
(\ref{resumz}) unphysical. Thus, the
parameterization (\ref{g1_2}) must be used only  at $x\leq 10^{-2}$.

\begin{figure}[t]
\epsfysize 10cm
\epsffile{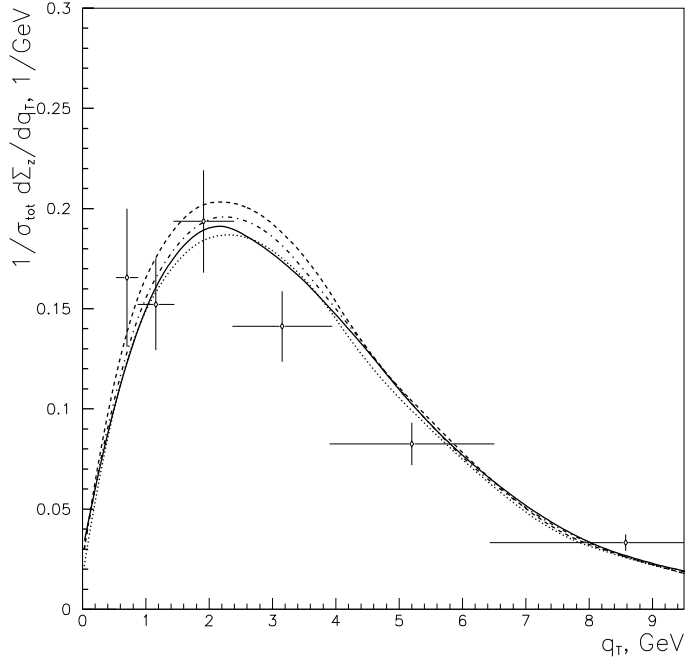}
\caption{\label{scale1} The dependence of the resummed $z$-flow
on the perturbative scale $\mu$ and the constants $C_1, C_2$. The Figure
corresponds to $\langle x\rangle =0.0049,\ \langle Q^2\rangle =32.6 \
\mbox{GeV}^2$.  The curves correspond to:  $C_1=2 e^{-\gamma}$, $C_2 =
e^{-3/4}, \mu = Q$ (solid); $C_1=2 e^{-\gamma}$, $C_2 = e^{-3/4}, \mu = 2
Q$ (dashed); $C_1=2 e^{-\gamma}$, $C_2 = e^{-3/4}, \mu = 0.5 Q$ (dotted);
$C_1=2 e^{-\gamma}$, $C_2 = 1, \mu = Q$ (dot-dashed).
}
\end{figure}

The theoretical results in Fig.~\ref{zqt} were obtained
using the
kinematic correction to the asymptotic and resummed cross-sections at
non-zero $q_T$, which was discussed in  Section \ref{5}.
As can be seen from Fig.~\ref{ixqtfl_dep}, without this correction the
agreement between the resummation calculation and the data is still good in
the region $q_T \leq 2\ \mbox{GeV} \leq Q$, where the resummation calculation
is truly applicable. The kinematic correction
improves the agreement between the resummed $z$-flow and
the data in the region   $q_T \geq 2 \ \mbox{GeV}$.
In this region, the
theoretical prediction without the kinematic correction significantly
exceeds the data. As  explained in  Section \ref{5}, this can be
attributed to differences between the phase space of the perturbative
and resummed pieces. In the case of the resummed
cross-section the phase space
may expand to much lower $x$ and $z$ than  is allowed
for the perturbative cross-section. Consequently, the
large
magnitude of PDFs and FFs at small $x$ and $z$ may spoil the
cancellation between the resummed and asymptotic piece at large $q_T$.
This difference
can be corrected by redefinition (\ref{tildexz})
of the variables $x$ and $z$
in the resummed and asymptotic
pieces.
\begin{figure}
\epsfysize 10cm
\epsffile{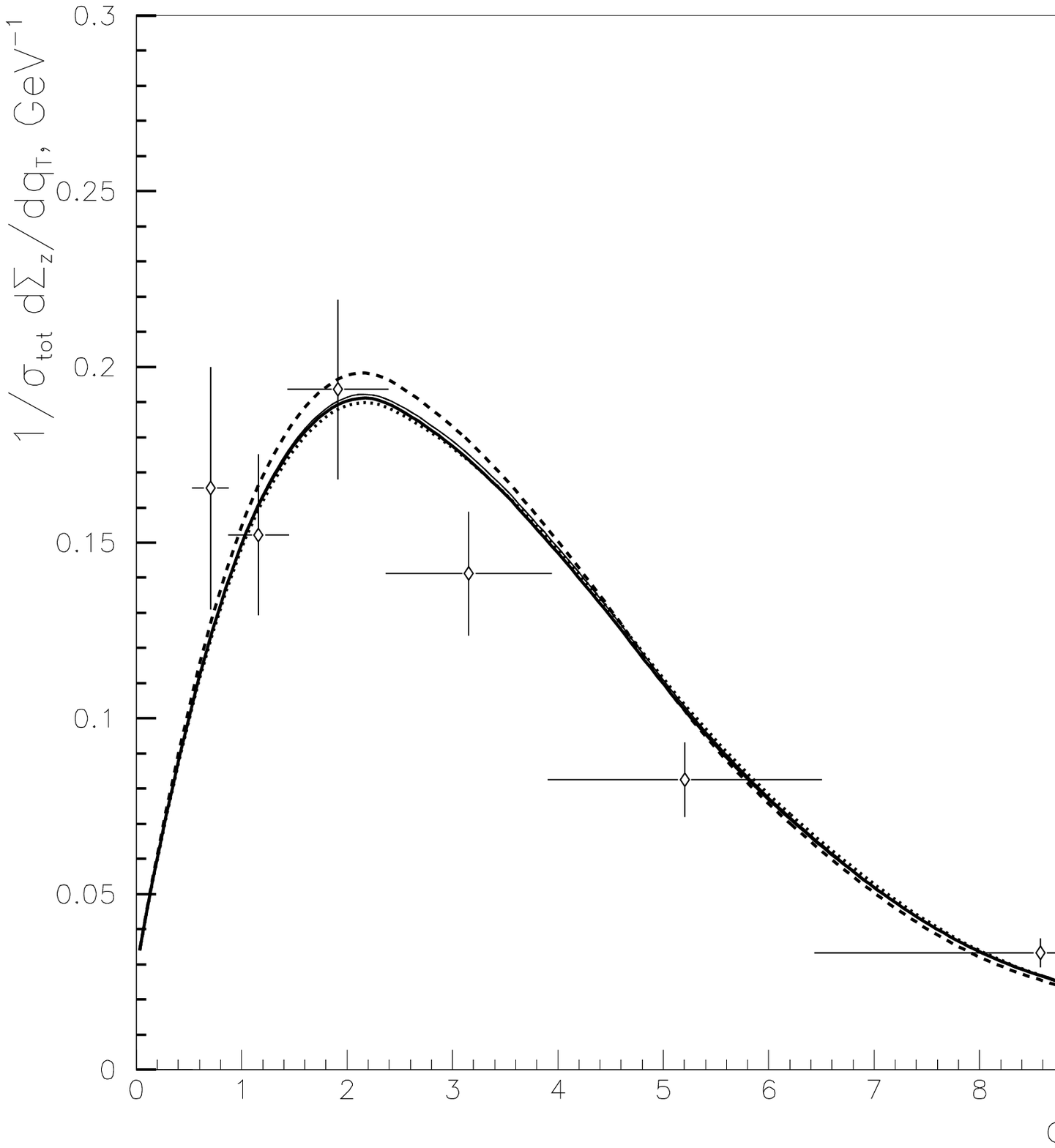}
\caption{\label{q0bmax} Dependence of the resummed $z$-flow on  $Q_0$
and
$b_{max}$. The curves correspond to
$Q_0 = 1\ \mbox{GeV}$, $b_{max}=1 \ \mbox{GeV}^{-1}$ (thick solid);
$Q_0 = 1\ \mbox{GeV}$, $b_{max}=0.5 \ \mbox{GeV}^{-1}$ (dashed);
$Q_0 = 1\ \mbox{GeV}$, $b_{max}=2 \ \mbox{GeV}^{-1}$ (dotted);
$Q_0 = 0.5\ \mbox{GeV}$, $ b_{max}=1 \ \mbox{GeV}^{-1}$ (thin solid);
$Q_0 = 2\ \mbox{GeV}$, $b_{max}=1 \ \mbox{GeV}^{-1}$ (dot-dashed).
The graph corresponds to
$\langle x\rangle =0.0049,\ \langle Q^2 \rangle = 32.6 \ \mbox{GeV}^2$.
}
\end{figure}

One of the advantages of the resummed cross-sections is that, by their
construction, they are less dependent on the choice of the
renormalization and factorization scales of the problem, and on
the end points $C_1/b$ and $C_2 Q$ in the integral of the perturbative
Sudakov factor (\ref{sud_p}). The reason is that the
scale variations in the perturbative part
of (\ref{resumz})
are compensated by the variation
of the term $g^{(2)}(b, Q)$
in the nonperturbative Sudakov factor (\ref{SzNP2}).

In Fig.~\ref{scale1} we show the
resummed $z$-flow for different choices of perturbative
renormalization and factorization scale $\mu$, varying between $0.5 Q$
and $ 2 Q$, and for a different choice of the constants
$C_1$ and $C_2$ (the ``canonical choice'' $C_1= 2 e^{-\gamma},\ C_2 =
1$). As expected, the resummed cross-section shows little variation
with the changes of $\mu, \ C_1, \ C_2$.

We have also checked the stability  of the resummed $z$-flow
under variation of  the momentum transfer scale $Q_0$ in the logarithm of
$g^{(2)}(b,Q)$, and under variation of
the parameter $b_{max}$ separating the
perturbative and nonperturbative dynamics in (\ref{resumz}).
Fig.~\ref{q0bmax} shows the $z$-flows for variations of $Q_0$ and
$b_{max}$
by factors of $1/2$ and $2$. As can be seen in
Fig.~\ref{q0bmax}, the variation
of the resummed $z$-flow is small. The stability of the resummed
$z$-flow under the variation of $b_{max}, Q_0, C_1, C_2$ indicates
that the choice of these parameters has less influence on the shape of
the resummed cross-section than the free parameters in $g^{(1)}(b,x)$.
It also illustrates the fact that the existing HERA
data is relatively insensitive  to the parameterization
of the $Q$-dependent function
$g^{(2)} (b, Q)$, which can be studied in a more detailed fashion
once the data in a larger range of $Q$ become available.

\begin{figure}[H]
\epsfysize 18cm
\epsffile{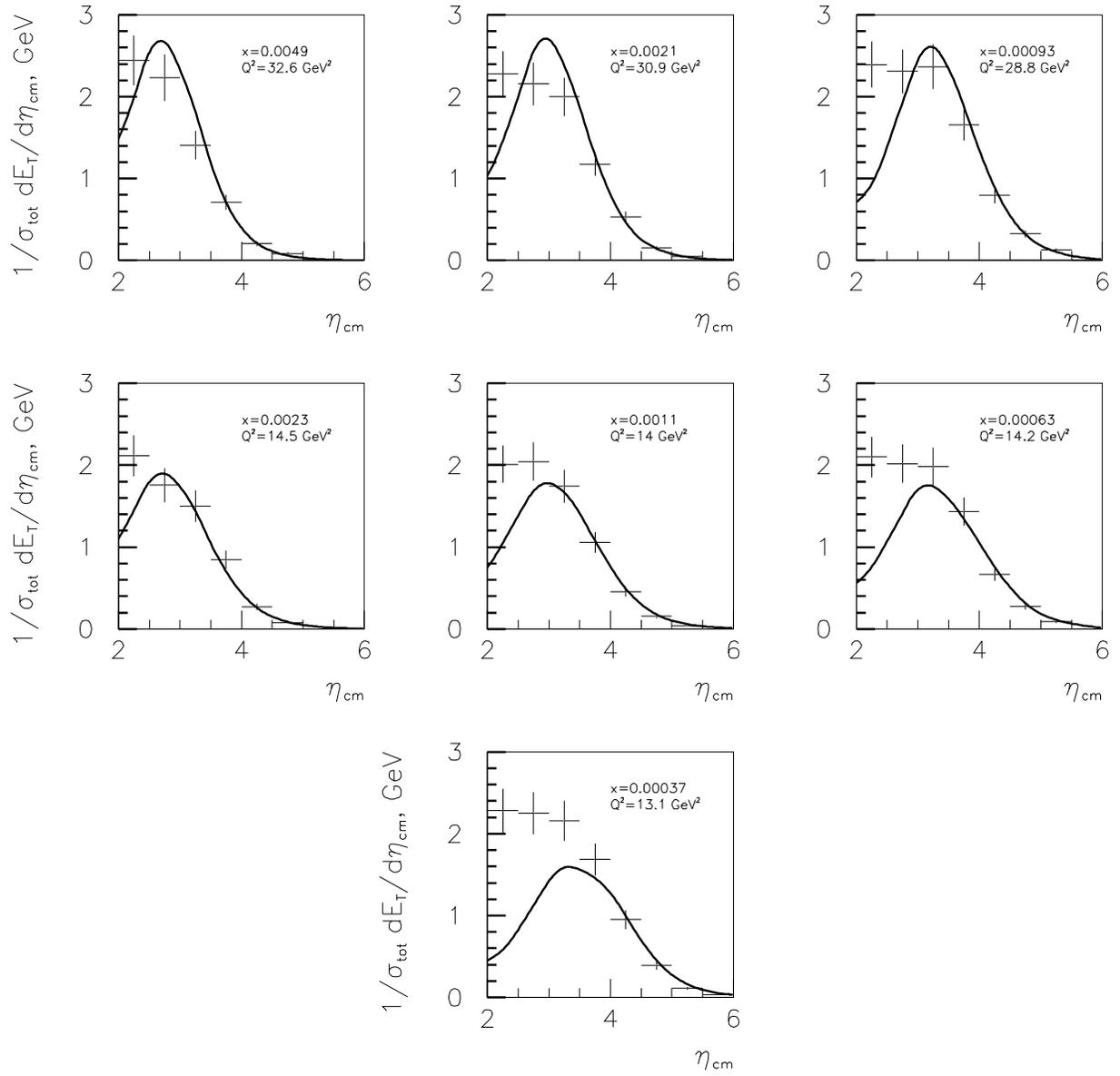}
\caption{\label{ETfig} The $\vg p$ \cms
pseudorapidity distributions of
the transverse energy flow in the current region. The data are from
{\protect\cite{H1z}}.
}
\end{figure}

Finally, in Fig.~\ref{ETfig} we  replot the results of our calculation
presented in Fig.~\ref{zqt} as the \cms    pseudorapidity distributions of
the transverse energy flow $\langle E_T \rangle$.
This quantity is obtained by
the transformation (\ref{Sz2ET}).  The small-$q_T$ region, where the
resummation formalism is valid, corresponds to large pseudorapidities.
In this region, the agreement between our calculation and the data is good.
At smaller pseudorapidities (larger $q_T$), one sees the above-mentioned
excess of the data over the perturbative NLO calculation. In the $\langle
E_T \rangle$ vs. $\eta_{cm}$ plot,  this excess is magnified because
of  the
factor $q_T^2$ in the transformation (\ref{Sz2ET}).

\newpage
\newpage
\begin{figure}[H]
\epsfxsize 15cm
\epsffile{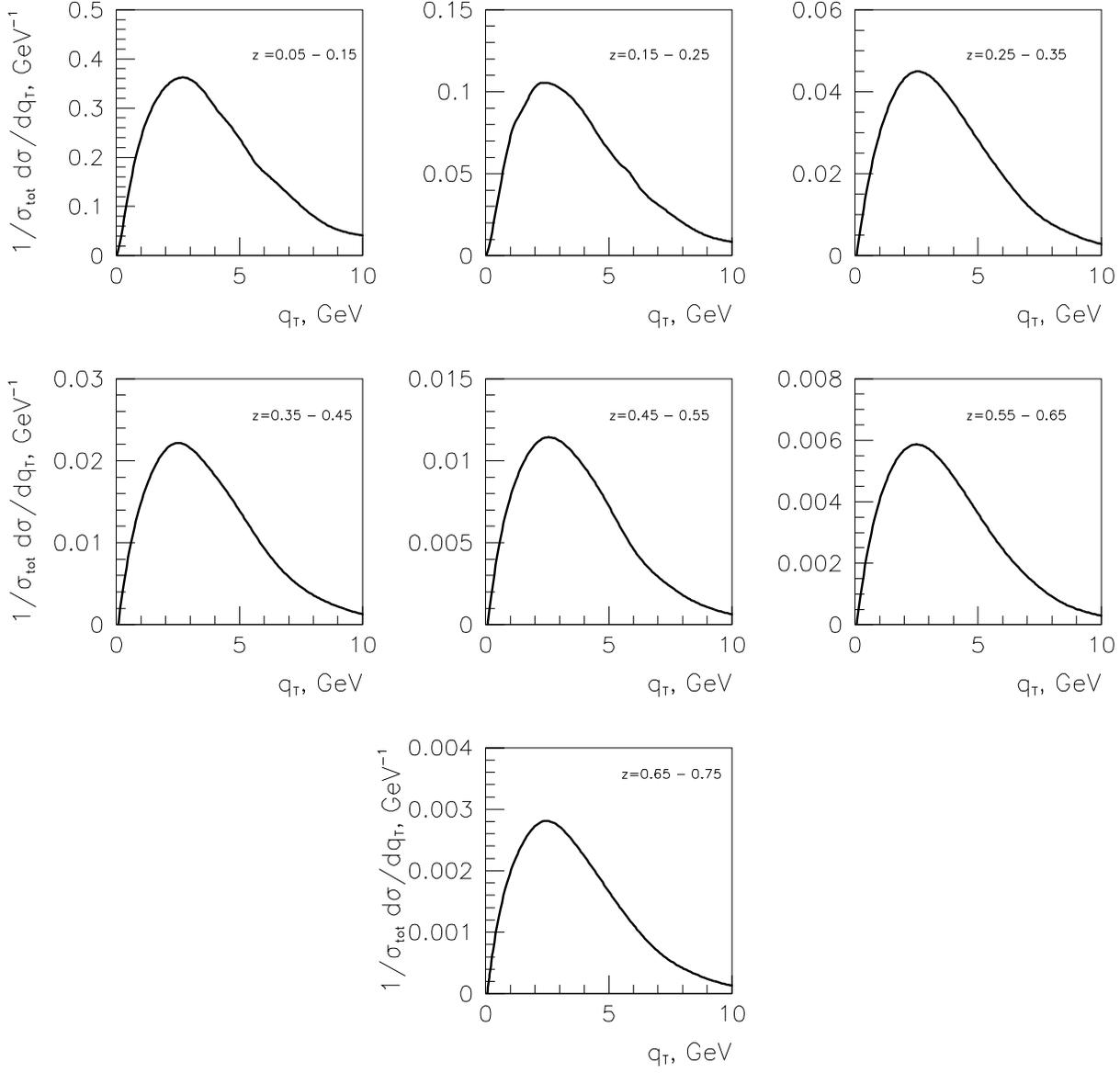}
\caption{\label{chgd} The charged particle multiplicity at
$\langle x\rangle =0.0044$, $\langle Q^2 \rangle = 35 \ \mbox{GeV}^2$.}
\end{figure}
\newpage

\subsection{Multiplicity of charged particle production}

The H1 and ZEUS
collaborations recently published the pseudorapidity
distributions of  charged
particle multiplicity at small values of $x$ and $Q^2$
\cite{HERAchgd}.
The data presented was organized in the same bins of $x$ and $Q^2$ as
those used in the H1 analysis of the transverse energy flow \cite{H1z}.
Also, the experiment E665 at Fermilab has provided  extensive data on
charged particle multiplicity covering the region of larger $x$
\cite{E665}.  However, a comparison of these data  with our calculation,
analogous to the one described in the previous subsection, is
complicated by several obstacles.

In comparison to  energy flows,  charged particle
multiplicity depends on the additional variable $z$ which controls the
fragmentation of a final-state parton into the observed
hadrons. Therefore the pQCD hadroproduction
cross-sections depend on parton fragmentation functions
$D (\xi_b,\mu)$,
which at present are known well only in the region $0.1 \leq
\xi_b \leq 0.8$ \cite{BKK}.

The
non-perturbative Sudakov factor (\ref{SzNP}) for particle
multiplicity can depend on $z$. Furthermore, in
principle the non-perturbative Sudakov factor can be different for
different types of  initial and final states (see the discussion of
Eq. (\ref{g2}) above). As a first approximation, in the analysis
below we will ignore this difference.

On the other hand, in the existing data
on charged particle pseudorapidity or transverse momentum distributions
\cite{HERAchgd,E665} the dependence on the final-state
fragmentation variable is not separated from the other
variables. Therefore more complicated fitting of the data is necessary
to disentangle the $q_T$ and $z$ dependences.

More importantly, our calculation was made under the assumption
that all the participating particles, including the final-state hadrons,
are massless. Because of this assumption, the production of soft
final-state hadrons, with $z=0$ is allowed. This contradicts the situation of
the experiments at HERA and Fermilab, in which there is a non-zero
 minimal value of $z$
determined by the finite mass of the observed hadron.
It follows from the definition (\ref{z}) of $z$ and the formulas
(\ref{pAc}, \ref{pBc}) for the initial and final hadron momenta in the $\vg
p$ \cms frame, that \beq z =\frac{p_{B+}^{cm}}{W}\geq \frac{m_B}{W},
\label{zmin}
\eeq
where \bdm p_{B+}^{cm} = E_B^{cm} + p_{B z}^{cm}
\mbox{ and }W^2 = Q^2 \Bigl( \frac{1}{x}-1 \Bigr).
\edm

Using (\ref{zmin}), one can show that the multiplicity
distributions presented in \cite{HERAchgd} receive contributions
from charged pions with $z \approx 10^{-3}$ or lower.
This means that for the experimental cuts used in the H1 analysis,
the multiplicity receives significant contribution from the region $z
\leq 0.1$ where the fragmentation functions are poorly known, and
mass effects are important.

In Fig.~\ref{chgd} we present the resummed multiplicity in the
 region $z\geq  0.1$, where the uncertainties
in our knowledge of the fragmentation functions and the mass effects
are minimal. The results in Fig.~\ref{chgd} are given for
 $x=0.0044, \ Q^2 = 35 \ \mbox{GeV}^2$ and various bins of
$z$. In the calculation, we used the CTEQ4M PDFs
\cite{CTEQ} and the fragmentation functions from \cite{BKK}. We show
the prediction using the $z$-independent non-perturbative Sudakov
factor (\ref{SzNP2}) obtained from the study of the energy flows.
We also assume that the charged particle production rate is dominated
by  fragmentation of partons into charged pions, kaons and
protons, and that the non-perturbative Sudakov factors for these
types of particles are the same.
In spite of many assumptions that go into this calculation, it will be
 interesting to compare it with the experimental data.

\section{Conclusions \label{7}}
In this paper, we have presented a formalism for
all-order resummation of large logarithms
arising in hadroproduction in
the current region of deep-inelastic scattering, {\it i.e.} for large
pseudorapidity of the final-state hadrons in the photon-proton 
cm.~frame. We
found that the formalism describes well  the behavior of the
transverse energy flows measured at HERA \cite{H1z}
in the region of large hadronic 
c.m.~pseudorapidity $\eta \geq 3.0$. At smaller pseudorapidities, we
found a deficit of the NLO rate compared to the existing
data. Evidently, this is a signature of the importance of the NNLO
corrections, which were not studied in this paper.

The formalism presented here can also be directly applied to the study of
hadron multiplicities. In this case, however, additional
care is required in treating the final-state fragmentation of
partons and the effects of the mass of the final-state hadron.
In view of this, reanalysis of charged particle production
multiplicity measured by  H1, ZEUS and Fermilab-E665, with the goal
to separate the $z$ and $q_T$ dependence, will be very useful to study
the properties of the non-perturbative part of the Sudakov factor,
and to improve the applicability of the perturbative calculation in
the current region.

\section*{Acknowledgements}
The authors would like to thank the CTEQ Collaboration,
C. Balazs, B. Harris,
M. Klasen, M. Kramer, C. Schmidt, J. Bartels, Wu-Ki Tung
for very helpful discussions. 
Some preliminary calculations were done by Kelly McGlynn.
This work was supported in part by the NSF
under grant PHY--9802564.

\section*
{Appendix A. Transformation from the hadron frame to the  HERA lab frame}
In this Appendix, we summarize the relationships between the
relativistic invariants used in this paper,
the hadron frame variables,
and the particle momenta in the HERA lab frame.

The definition of the HERA lab frame  is that the proton ($A$)
moves in the $+z$ direction, with energy $E_{A}$, and the
incoming lepton moves in the $-z$ direction with energy $E$.
The momenta of the incident particles are
\begin{equation}
p_A^{\mu}=(E_{A},0,0,E_{A}),
\end{equation}
\begin{equation}
l^{\mu}=(E,0,0,-E)~.
\end{equation}

The outgoing lepton has energy $E^{\prime}$ and scattering angle
$\theta$ relative to the $-z$ direction.
We define the $x$-axis of the HERA frame such that the outgoing lepton
is in the $xz$-plane; that is,
\begin{equation}
l^{\prime\mu}_l=(E^{\prime},-E^{\prime}\sin\theta,0,
-E^{\prime}\cos\theta)~.
\end{equation}
The observed hadron ($B$) has energy $E_{B}$ and scattering angle
$\theta_{B}$ with respect to the $+z$ direction, and azimuthal
angle $\phi_{B}$; thus its momentum is
\begin{equation}
{p_{B}^{\mu}}_l=(E_{B},E_{B}\sin\theta_{B}\cos\phi_{B},
E_{B}\sin\theta_{B}\sin\phi_{B},E_{B}\cos\theta_{B})~.
\label{pTBl}
\end{equation}

The scalars $x$ and $Q^{2}$ are completely determined by measuring the
energy and the scattering angle of the outgoing lepton,
\begin{equation}
Q^{2}=2 E E^{\prime}(1-\cos\theta),
\end{equation}
\begin{equation}
x=\frac{E E^{\prime}(1-\cos\theta)}{E_{A}
\left[2E-E^{\prime}(1+\cos\theta)\right]}~.
\end{equation}
The scalars $z$ and $q_{T}^{2}$ depend on the outgoing hadron
and lepton as
\begin{equation}
z=\frac{\beta E_{B}(1-\cos\theta_{B})}{Q},
\end{equation}
\beq
q_T^2= \frac{2\,{Q^2}\,
     \left( 1 -
       \left( y -1  \right)
          \,{{\beta }^2}
        \right) }{{{\beta }^
      2 (1- \cos \theta_B)}}
\Biggl[ \sin^2 \frac{\theta_B - \theta_*}{2} +
\sin \theta_B \sin\theta_* \sin^2 \frac{\phi_B}{2} \Biggr],
\eeq
where
\beq
y =\frac{Q^2}{x \SeA}, \ 
\beta = \frac{ 2 x E_A}{Q},\ 
 cot \frac{\theta_*}{2} = \beta \sqrt{1-y}.
\eeq

The 
angle $\phi$ and the boost parameter $\psi$, which are used for the
angular decomposition of the cross-sections in the hadron Breit frame, 
can be found as
\beq
\cos \phi = \frac{Q}{2\,{\sqrt{1 - y}}\,
     {q_T}}\,\left[ 1 - y -
       \frac{\cot^2 (\theta_B/2)}{\beta^2} +
       \frac{q_T^2}{Q^2} \right],
\eeq
\beq
 \cosh \psi =\frac{2}{y} -1.
\eeq

\section*{Appendix B. The perturbative cross-section and 
$z$-flow distribution}
In this Appendix, we collect the formulas for the NLO parton level
cross-sections $d\hat \sigma_{ba} / (d \xh d\zh d Q^2 d q_T^2 d\phi)$.

According to  (\ref{hadcs}), the hadron level cross-section
$d\sigma_{BA} /( d x d z d Q^2 d q_T^2 d\phi)$ is related to the parton
cross-sections $d\sigma_{ba} /( d \xh d \zh d Q^2 d q_T^2 d\phi)$ as
\bdm
\frac{d\sigma_{BA}}{dx dz dQ^{2} dq_{T}^{2} d\phi}=\sum_{a,b}
\int_{z}^{1}
\frac{d\xi_{b}}{\xi_{b}} D_{B/b}(\xi_{b})\int_{x}^{1}
\frac{d\xi_{a}}{\xi_{a}}
F_{a/A}(\xi_{a})
\frac{d\hat{\sigma_{ba}}}{d\hat{x} d\hat{z} dQ^{2} dq_{T}^{2} d\phi}.
\edm

At non-zero $q_T$, the parton cross-section receives the contribution from
the real emission diagrams (Fig.~\ref{diags}e-f); it can be expressed as
\bea
&& \frac{d \hat \sigma_{ba}}{d \xh d \zh d Q^2 d q_T^2 d\phi} =
\nonumber \\
&&\frac{ \sigma_0 F_l}{4 \pi \SeA Q^2} \frac{\alpha_s}{\pi}
\delta \Biggl[ \frac{q_T^2}{Q^2}- \Bigl(\frac{1}{\xh}-1\Bigr)
\Bigl(\frac{1}{\zh}-1\Bigr)
\Biggr]
\sum_j e^2_j \sum_{\alpha=1}^{4} f^{(\alpha)}_{ba}(\xh, \zh, Q^2, q_T^2)
A_{\alpha}(\psi, \phi),
\label{sighat}
\eea
with the same notations as in Section \ref{3}.
In this formula,
\bea
&&\sum_{\alpha=1}^{4} f^{(\alpha)}_{jk}(\xh, \zh, Q^2, q_T^2)
A_{\alpha}(\psi, \phi)  = 2 \delta_{jk} C_F \xh \zh
\Biggl\{
\Bigl[\frac{1}{q_T^2}\biggl(\frac{Q^4}{\xh^2 \zh^2} + (Q^2 - q_T^2)^2
\biggr)+ 6 Q^2 \Bigr] A_1
\nonumber \\
&+& 2 Q^2 (2 A_2 + A_4) + 2 \frac{Q}{q_T} (Q^2 +q_T^2) A_3
\Biggr\} ;
\eea
\bea
& &\sum_{\alpha=1}^{4} f^{(\alpha)}_{jg}(\xh, \zh, Q^2, q_T^2)
A_{\alpha}(\psi, \phi)  =
\nonumber\\
&&\xh (1-\xh)
\Biggl\{
\Bigl[\frac{Q^4}{q_T^2}
\biggl(\frac{1}{\xh^2 \zh^2}- \frac{2}{\xh \zh} + 2 \biggr)
+ 2 Q^2 (5  -\frac{1}{\xh} -\frac{1}{\zh}) \Bigr] A_1
\nonumber\\
&+&4 Q^2 (2 A_2 + A_4) + 2 \frac{Q}{q_T}
( 2 (Q^2 +q_T^2)-\frac{Q^2}{\xh \zh}) A_3
\Biggr\} ;
\eea
\bea
& &\sum_{\alpha=1}^{4} f^{(\alpha)}_{gj}(\xh, \zh, Q^2, q_T^2)
A_{\alpha}(\psi, \phi)  =
\nonumber \\
& &2 C_F \xh (1-\zh)\Biggl\{
\Bigl[\frac{1}{\tilde q_T^2}\biggl(\frac{Q^4}{\xh^2 (1-\zh)^2} +
(Q^2 - \tilde q_T^2)^2
\biggr)+ 6 Q^2 \Bigr] A_1
\nonumber \\
&+& 2 Q^2 (2 A_2 + A_4) + 2 \frac{Q}{\tilde q_T} (Q^2 +\tilde q_T^2) A_3
\Biggr\}. \label{111}
\eea
In (\ref{111}),
\beq
\tilde q_T = \frac{\zh q_T}{1-\zh}.
\eeq
The index $j$ corresponds to a quark (antiquark) of a type $j$, the
index $g$ corresponds to a gluon.

From (\ref{hadcs}), it is possible to derive the perturbative $z$-flow 
distribution, 
\bdm
\frac{d \Sigma_z}{d x d Q^2  d q_T^2d \phi}
\equiv \sum_B \int_{z_{min}}^1 z d z \frac{d \sigma_{BA}}{ d x d z d
Q^2 d q_T^2 d\phi} =
\edm
\beq
\frac{ \sigma_0 F_l}{4 \pi \SeA Q^2} \frac{\alpha_s}{\pi}
\sum_{a,b}\sum_j e^2_j \int_{x}^{1}
\frac{d\xi_{a}}{\xi_{a}-x}
F_{a/A}(\xi_{a})
\zh^3 \xh
\sum_{\alpha=1}^{4} f^{(\alpha)}_{ba}(\xh, \zh, Q^2, q_T^2)
A_{\alpha}(\psi, \phi).
\label{zx}
\eeq
It depends on the same functions $f_{ba} (\xh, \zh, Q^2, q_T^2)$, with
the parton variable $\zh$ determined by the $\delta$-function in
(\ref{sighat}),
\beq
\zh=\frac{1-\xh}{(q_T^2/Q^2-1) \xh +1}.
\eeq

\end{document}